\documentclass[a4paper,11pt]{article}
\pdfoutput=1

\usepackage{jheppub}
\usepackage{amssymb}
\usepackage{graphicx}
\usepackage[dvipsnames,table]{xcolor}
\usepackage{slashed}
\usepackage{hyperref} 
\usepackage{xspace}
\usepackage[tight]{subfigure}
\usepackage{amsmath}
\usepackage{amsfonts}
\usepackage{mathrsfs}
\usepackage{comment}
\usepackage{afterpage}
\usepackage{verbatim}
\usepackage{booktabs}
\usepackage{array}
\usepackage[title,titletoc]{appendix}
\usepackage{tabularx}

\newcolumntype{C}[1]{>{\centering\arraybackslash}p{#1}}

\newcommand{\nnb}{\nonumber}
\newcommand{\mc}{\mathcal}

\def\refeq#1{\mbox{Eq.~\eqref{#1}}}

\def\reffi#1{\mbox{Figure~\ref{#1}}}

\def\refta#1{\mbox{Table~\ref{#1}}}

\def\refse#1{\mbox{Section~\ref{#1}}}
\def\refses#1{\mbox{Sections~\ref{#1}}}

\def\citere#1{\mbox{Ref.~\cite{#1}}}
\def\citeres#1{\mbox{Refs.~\cite{#1}}}

\newcommand{\newc}{\newcommand}
\newc{\beq}{\begin{equation}}
\newc{\eeq}{\end{equation}}
\newc{\beqn}{\begin{eqnarray}}
\newc{\eeqn}{\end{eqnarray}}
\newc{\bit}{\begin{itemize}}
\newc{\eit}{\end{itemize}}
\newc{\ben}{\begin{enumerate}}
\newc{\een}{\end{enumerate}}
\newc{\bce}{\begin{center}}
\newc{\ece}{\end{center}}
\newc{\bfi}{\begin{figure}}
\newc{\efi}{\end{figure}}

\newcommand{\cM}{\ensuremath{\mathcal{M}}\xspace}

\newcommand{\ri}{\mathrm i}

\newcommand{\rT}{{\mathrm{T}}}

\newcommand{\rL}{{\mathrm{L}}}

\newcommand{\rU}{{\mathrm{U}}}

\newcommand{\ie}{{i.e.}\ }
\newcommand{\eg}{{e.g.}\ }

\newcommand{\GeV}{\ensuremath{\,\text{GeV}}\xspace}
\newcommand{\TeV}{\ensuremath{\,\text{TeV}}\xspace}

\newcommand{\PH}{\ensuremath{\text{H}}\xspace}
\newcommand{\Pj}{\ensuremath{\text{j}}\xspace}
\newcommand{\Pp}{\ensuremath{\text{p}}}
\newcommand{\Pe}{\ensuremath{\text{e}}\xspace}

\newcommand{\Pt}{\ensuremath{\text{t}}\xspace}

\newcommand{\PW}{\ensuremath{\text{W}}\xspace}
\newcommand{\PZ}{\ensuremath{\text{Z}}\xspace}

\newcommand{\pt}[1]{p_{\rT,{#1}}}

\newcommand{\Mt}{\ensuremath{m_\Pt}\xspace}
\newcommand{\MH}{\ensuremath{M_\PH}\xspace}
\newcommand{\MWOS}{\ensuremath{M_\PW^\text{OS}}\xspace}
\newcommand{\MW}{\ensuremath{M_\PW}\xspace}
\newcommand{\MZOS}{\ensuremath{M_\PZ^\text{OS}}\xspace}
\newcommand{\MZ}{\ensuremath{M_\PZ}\xspace}

\newcommand{\Gt}{\ensuremath{\Gamma_\Pt}\xspace}
\newcommand{\GH}{\ensuremath{\Gamma_\PH}\xspace}
\newcommand{\GZ}{\ensuremath{\Gamma_\PZ}\xspace}
\newcommand{\GZOS}{\ensuremath{\Gamma_\PZ^\text{OS}}\xspace}
\newcommand{\GW}{\ensuremath{\Gamma_\PW}\xspace}
\newcommand{\GWOS}{\ensuremath{\Gamma_\PW^\text{OS}}\xspace}

\newcommand{\GF}{\ensuremath{G_\mu}}
\newcommand{\alphas}{\ensuremath{\alpha_\text{s}}\xspace}

\newcommand{\order}[1]{\ensuremath{\mathcal{O}{\left(#1\right)}}\xspace}

\newcommand{\recola}{{\sc Recola}\xspace}

\newcommand{\mocanlo}{{\sc MoCaNLO}\xspace}
\newcommand{\bbmc}{{\sc BBMC}\xspace}
\newcommand{\collier}{{\sc Collier}\xspace}

\newcolumntype{.}{D{.}{.}{-1}}
\newcolumntype{d}[1]{D{.}{.}{#1}}
\colorlet{tableoverheadcolor}{gray!37.5}
\colorlet{tableheadcolor}{gray!25}
\colorlet{tablerowcolor}{gray!12.5}

\def\draftdate{\relax}
\def\mda{\relax}
\def\mua{\relax}
\def\mla{\relax}
\def\draft{
\def\thtystars{******************************}
\def\sixtystars{\thtystars\thtystars}
\typeout{}
\typeout{\sixtystars**}
\typeout{* Draft mode!
         For final version remove \protect\draft\space in source file *}
\typeout{\sixtystars**}
\typeout{}
\def\draftdate{\today}
\def\mua{\marginpar[\boldmath\hfil$\uparrow$]%
                   {\boldmath$\uparrow$\hfil}\color{black}%
                    \typeout{marginpar: $\uparrow$}\ignorespaces}
\def\mda{\color{red}\marginpar[\boldmath\hfil$\downarrow$]%
                   {\boldmath$\downarrow$\hfil}%
                    \typeout{marginpar: $\downarrow$}\ignorespaces}
\def\mla{\marginpar[\boldmath\hfil$\rightarrow$]%
                   {\boldmath$\leftarrow $\hfil}%
                    \typeout{marginpar: $\leftrightarrow$}\ignorespaces}
\def\Mua{\marginpar[\boldmath\hfil$\Uparrow$]%
                   {\boldmath$\Uparrow$\hfil}\color{black}%
                    \typeout{marginpar: $\uparrow$}\ignorespaces}
\def\Mda{\color{red}\marginpar[\boldmath\hfil$\Downarrow$]%
                   {\boldmath$\Downarrow$\hfil}%
                    \typeout{marginpar: $\downarrow$}\ignorespaces}
\def\Mla{\marginpar[\boldmath\hfil\textcolor{red}{$\Rightarrow$}]%
                   {\boldmath\textcolor{red}{$\Leftarrow $}\hfil}%
                    \typeout{marginpar: $\leftrightarrow$}\ignorespaces}
\overfullrule 5pt
\oddsidemargin 15mm
\marginparwidth 29mm
}

\title{Electroweak corrections to doubly polarised $\PW\PZ$ scattering at the LHC}
\author[a]{Ansgar Denner,}
\author[a]{Robert Franken,}
\author[a]{Christoph Haitz,}
\author[a,b]{Daniele Lombardi,}
\author[c]{Giovanni Pelliccioli} 
\affiliation[a]{Institut f\"ur Theoretische Physik und Astrophysik,
  Universit\"at W\"urzburg, \\Emil-Hilb-Weg~22, 97074 W\"urzburg, Germany}
\affiliation[b]{Dipartimento di Fisica, Universit\`a  di Torino, and INFN, Sezione di Torino, Via P. Giuria 1,
I-10125 Torino, Italy}
\affiliation[c]{Universit\`a degli Studi di Milano--Bicocca
  and INFN Sezione di Milano--Bicocca,
  Piazza della Scienza 3, 20126 Milano, Italy}
\emailAdd{ansgar.denner@uni-wuerzburg.de}
\emailAdd{robert.franken@uni-wuerzburg.de}
\emailAdd{christoph.haitz@uni-wuerzburg.de}
\emailAdd{daniele.lombardi@unito.it}
\emailAdd{giovanni.pelliccioli@unimib.it}

\abstract{
  We present a calculation of next-to-leading-order electroweak corrections
  to the vector-boson scattering (VBS) process resulting in
  leptonically decaying $\PW$ and $\PZ$~bosons
  in association with two jets at the LHC.
  The VBS process is computed for both polarised and unpolarised intermediate bosons,
  exploiting the pole approximation and the separation of helicity states in tree-level and one-loop
  amplitudes.
  A phenomenological analysis is carried out for a realistic fiducial setup
  at a 13.6\TeV LHC collision energy, highlighting different patterns for the various
  polarisation states both at integrated and at differential level.
  This study provides theoretical predictions that are necessary
  to perform a sound characterisation of the spin structure of VBS processes
  with full LHC data.
}%

\preprint{COMETA-2025-49}

\begin{document}
\maketitle
\flushbottom

\section{Introduction}\label{sec:intro}
The scattering of electroweak (EW) gauge bosons constitutes an optimal process to probe
the mechanism of electroweak symmetry breaking (EWSB).
Such a process not only offers a stringent test of the EW~gauge and scalar sectors of
the Standard Model (SM),
but also serves as a handle to access potential effects that originate from dynamics beyond the SM.
The high-energy scattering of longitudinally polarised $\PW$ and $\PZ$ bosons is governed in the SM by
delicate cancellations between pure-gauge-boson contributions and Higgs-mediated ones, both of which
would separately violate perturbative unitarity. Such cancellations can be spoiled by
new-physics effects that change the SM EWSB pattern, leading to unitarity restoration
at higher energy scales than in the SM.

Since EW bosons are unstable, accessing their properties at colliders is only possible
via their decay products (leptons, neutrinos, and jets). Although intricate,
both from an experimental and from a theoretical viewpoint, measuring the polarisation state of
EW bosons is one of the main targets of the LHC physics programme. In particular, ATLAS and CMS
have already measured polarised-boson cross sections and polarisation fractions
through {\it polarised-template fits} in inclusive
$\PW\PZ$ and $\PZ\PZ$ production \cite{Aaboud:2019gxl,CMS:2021icx,ATLAS:2022oge,ATLAS:2023zrv,ATLAS:2024qbd}
and in same-sign $\PW\PW$ scattering \cite{CMS:2020etf,ATLAS:2025wuw}.
This approach is substantially different from a plain analysis of
angular coefficients and relies
on a specific theory prediction (in the SM or beyond) for each polarisation state of the EW bosons
involved in a process.

The modelling of polarised bosons as intermediate states in LHC processes
has undergone a terrific improvement in recent years, with the paradigm being the
separation of polarisation states at the level of amplitudes and the exploitation
of the pole approximation \cite{Ballestrero:2017bxn,Ballestrero:2019qoy,Ballestrero:2020qgv,Denner:2020bcz,Denner:2020eck,Poncelet:2021jmj,Denner:2021csi,Le:2022lrp,Le:2022ppa,Denner:2022riz,Dao:2023pkl,Pelliccioli:2023zpd,Denner:2023ehn,Dao:2023kwc,Denner:2024tlu,Dao:2024ffg,Haisch:2025jqr} or the narrow-width approximation \cite{BuarqueFranzosi:2019boy,Pellen:2021vpi,Pellen:2022fom,Hoppe:2023uux,Javurkova:2024bwa}
to treat the bosons in a gauge-invariant manner.

While the state-of-the-art polarised-boson predictions for inclusive
boson-pair production have reached very high precision and accuracy
\cite{Denner:2020bcz,Denner:2020eck,Poncelet:2021jmj,Denner:2021csi,Le:2022lrp,Le:2022ppa,Denner:2022riz,Dao:2023pkl,Pelliccioli:2023zpd,Denner:2023ehn,Dao:2023kwc,Dao:2024ffg,Haisch:2025jqr,Hoppe:2023uux,Javurkova:2024bwa},
the description of VBS processes is currently limited to leading order (LO) matched to parton shower (PS)
in the {\sc Phantom} and {\sc MadGraph5\_aMC@NLO} event generators
\cite{Ballestrero:2017bxn,Ballestrero:2019qoy,BuarqueFranzosi:2019boy,Ballestrero:2020qgv},
with the exception of doubly polarised $\PW^+\PW^+$ scattering, which has been recently computed
at next-to-leading order (NLO) in the EW and the QCD couplings \cite{Denner:2024tlu}.
The most recent version of the {\sc Sherpa} general purpose generator \cite{Hoppe:2023uux,Sherpa:2024mfk}
is potentially capable of calculating arbitrary polarised multi-boson processes at approximate NLO QCD accuracy
including multi-jet merging and matching to PS, but a concrete application to VBS has not been
carried out yet.
In VBS processes, it is especially important to include at least NLO EW corrections,
which are known to be large and negative in typical fiducial LHC setups
\cite{Biedermann:2016yds,Denner:2019tmn,Denner:2020zit,Denner:2022pwc},
owing to logarithmic enhancements of virtual origin \cite{Denner:2000jv,Denner:2001gw}.
In this work we perform one step further in this direction, computing the
NLO EW corrections to doubly polarised $\PW\PZ$ scattering at the LHC.

The $\PW\PZ$ scattering has been measured with Run-2 data
\cite{ATLAS:2018mxa,CMS:2019uys,CMS:2020gfh,ATLAS:2024ini}
in the fully leptonic decay channel, but so far no results have been obtained
for its polarisation structure. Owing to the presence of a single neutrino, the $\PW\PZ$ scattering
allows for complete reconstruction of the final state, thanks to the exploitation of
neutrino-reconstruction techniques. This aspect, in spite of a lower signal rate and larger background
compared to same-sign $\PW\PW$ scattering, 
makes the $\PW\PZ$ VBS channel appealing from the experimental viewpoint,
especially in light of the data accumulation of the Run-3 and High-Luminosity LHC stages.
The SM modelling of the $\PW\PZ$ VBS signal is known up to NLO QCD \cite{Bozzi:2007ur} and NLO EW accuracy \cite{Denner:2019tmn}. The PS matching has been achieved at NLO QCD \cite{Jager:2018cyo,Jager:2024sij}. New-physics effects on this process have been studied in the context of effective field theories \cite{Delgado:2017cls,Garcia-Garcia:2019oig,Herrero:2021iqt,Bellan:2021dcy}.
The $\PW\PZ$ VBS process is also of interest to probe possible beyond-the-SM scalar sectors leading to charged-Higgs bosons, like two-Higgs-doublet \cite{Lee:1973iz,Glashow:1976nt,Paschos:1976ay,Branco:2011iw} or Higgs-triplet models \cite{Georgi:1985nv,Gunion:1989ci,Englert:2013zpa}.
The polarised signals in $\PW\PZ$ scattering have been computed in the pole
approximation at LO both in the SM and in the presence of an additional real Higgs singlet \cite{Ballestrero:2019qoy}. The NLO corrections and PS matching are still missing in the literature.

This paper aims at achieving exact NLO EW accuracy in the description of
$\PW\PZ$ scattering at the LHC with both bosons in a definite polarisation state.
It is organised as follows: In \refse{sec:calc} we show the details of our calculation, including a discussion of generic aspects of the scattering sub-process and more technical details regarding the definition of polarised signals in the pole approximation. A phenomenological analysis in a realistic LHC Run-3 scenario
is carried out in \refse{sec:results}, where we show both integrated and differential results for doubly polarised signals. Our conclusions are drawn in \refse{sec:conc}.
\vfill

\section{Details of the calculation}\label{sec:calc}
We consider the production of three charged leptons, two jets, and
missing transverse energy at the LHC,
\beq\label{eq:procdef}
\Pp\Pp\rightarrow \Pe^+\nu_{\Pe}\mu^+\mu^-\Pj\Pj\,,
\eeq
at the EW tree~level, \ie $\mc O(\alpha^6)$, and at NLO EW accuracy, \ie including
NLO corrections of $\mc O(\alpha^7)$.
Although at these pure EW perturbative orders the process receives contributions
from quark--quark-, quark--photon-, and photon--photon-induced partonic channels, we only consider
the first ones in this work, which are expected to be the dominant contributions. 
We have, however, checked that the photon-induced
channels contribute only at the one-percent level to
the unpolarised fiducial cross section in the double-pole approximation (DPA).
Additionally, the photon-induced
channels typically enhance the LO-suppressed triple-boson topologies, owing to the
potential presence of three jets in the final state. In this sense, the photon-induced
process can be regarded as part of the irreducible triple-boson background to
VBS.

While we compute also the full off-shell process defined in \refeq{eq:procdef}, the
focus of this work is on the {\it on-shell\/} signals with either polarised or unpolarised intermediate EW bosons, namely,
\beq
\label{eq:procdefDPA}
\Pp\Pp\rightarrow \PW^+_{\lambda}(\rightarrow\Pe^+\nu_{\Pe})\,\PZ_{\lambda'}(\rightarrow \mu^+\mu^-)\Pj\Pj\,,\qquad \lambda,\lambda'=\rL, \rT, \rU\,,
\eeq
with the labels $\rL,\,\rT$, and $\rU$ indicating a longitudinal, transverse, and unpolarised EW boson.
Before providing the definition of the (un)polarised signals in \refeq{eq:procdefDPA}, we recall
some general aspects of VBS that are relevant for our phenomenological study.

\subsection[General aspects of WZ scattering]{General aspects of $\PW\PZ$ scattering}\label{sec:2to2}
\begin{figure}
  \centering
  \includegraphics[width=0.9\textwidth]{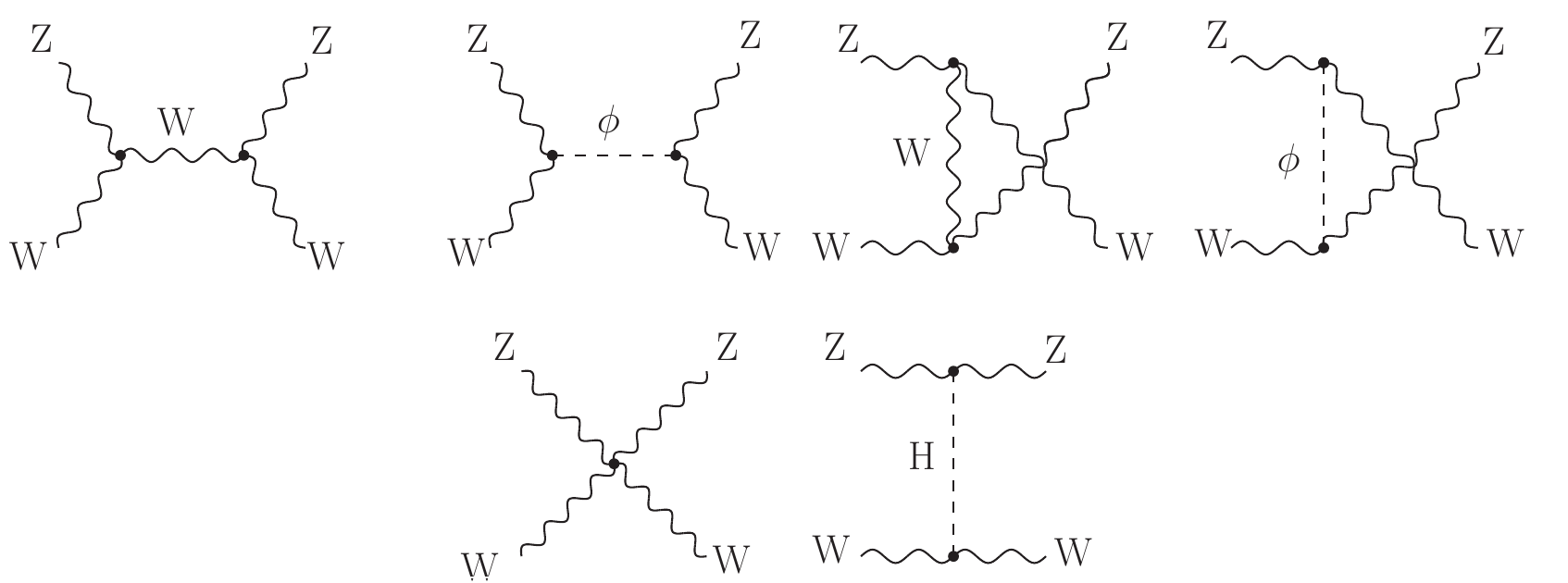}
  \caption{ Tree-level SM diagrams contributing to the $\PW^+\PZ\rightarrow\PW^+\PZ$ scattering process.
  }\label{fig:diags}
\end{figure}

It is a well-established result
\cite{Veltman:1989vw,Dawson:1989up,Passarino:1990hk,Valencia:1990jp,Denner:1997kq}
that in the $2\rightarrow 2$ scattering of longitudinally polarised bosons at high energy,
the unitarity violation owing to massive-gauge-boson diagrams
is cured by the inclusion of Higgs-exchange contributions,
thereby restoring unitarity in the LO SM amplitude.
Therefore, the physical amplitude for this scattering process receives contributions both from pure-gauge-boson diagrams, depending on the triple and quartic gauge couplings,
and from diagrams mediated by the Higgs boson, thus depending on the Higgs--gauge-boson couplings. The tree-level diagrams are depicted in \reffi{fig:diags}.
If all external, \ie on-shell, bosons are longitudinally polarised, in
the high-energy regime, $s\gg\MZ^2$, the contributions to the scattering
amplitude read \cite{Schwartz:2014sze},
\begin{align}
  \cM_{\rm qgc} ={}
  & g^2_{\rm w}\left(\frac{s}{4\MZ^2}\right)^2\left(3+6\cos\theta-\cos^2\theta\right)\nnb\\
  & -g^2_{\rm w}\left(\frac{s}{4\MZ^2}\right)^{\,}\left(1+\frac{\MW^2}{\MZ^2}\right)\left(1+3\cos\theta\right) + \mc O(1)
  \,,\label{eq:4v}\\
  \cM_{s+u,\,\PW\!/\!\phi} ={}
  & {-g^2_{\rm w}}\left(\frac{s}{4\MZ^2}\right)^2\left(3+6\cos\theta-\cos^2\theta\right)\nnb\\
  &  +g^2_{\rm w}\left(\frac{s}{4\MZ^2}\right)^{\,}\left[  \left(1+\frac{\MW^2}{\MZ^2}\right)
    \left(1+3\cos\theta\right)
-\left(\frac{\MZ^2}{2\MW^2}\right)\left(1-\cos\theta\right)
  \right] + \mc O(1)\,,\label{eq:3v}\\
  \cM_{t, \PH} ={}&
  g^2_{\rm w}
  \left(\frac{s}{4\MZ^2}\right)^{\,}\left(\frac{\MZ^2}{2\MW^2}\right)\left(1-\cos\theta\right)+\mc O(1)
  \,,\label{eq:h}
\end{align}
depending on the Mandelstam variable $s$ (the square
of the centre-of-mass (CM) energy)
and the scattering angle $\theta$. 
The parameters $\MW$ and $\MZ$ refer to the W- and Z-boson masses, respectively,
while $g_{\rm w}$ is the weak [SU(2)] gauge coupling.
Terms that are not enhanced by a factor $s/\MZ^2$ are implicitly
accounted for in the $\mc O(1)$ terms. 
The quartic-gauge-coupling diagrams contribute to  \refeq{eq:4v}, while the sum of triple-gauge-coupling terms with the
corresponding Goldstone-boson ones leads to \refeq{eq:3v}. For both $\cM_{s+u,\,\PW\!/\!\phi}$ and $\cM_{\rm qgc}$ the
unitarity-violating behaviour is individually quadratic in $s$, but gets linear in their sum, \ie $\cM_{\rm qgc}+\cM_{s+u,\PW\!/\!\phi} \sim s/\MZ^2$.
The very same linear growth with  $s$ (but with opposite sign) is
found for the Higgs-mediated contribution in \refeq{eq:h}, which therefore unitarises the SM amplitude.
It is then clear that the cancellation pattern depends on the details of the EWSB realisation in the SM.
In particular, given the expression for the vacuum expectation value of the Higgs field $v=2\MW/g_{\rm w}$,
a Higgs-less EW theory would violate perturbative unitarity at the
scale $\Lambda_{\rm ew} = \sqrt{8\pi}v\approx 1\TeV$, 
which can be therefore regarded as an upper bound on the EWSB scale \cite{Marciano:1989ns}. Although the SM realisation of EWSB is increasingly supported by experimental LHC data, neither extended Higgs sectors nor strongly interacting models are ruled out
\cite{ParticleDataGroup:2024cfk}. While the former class of models is more minimal and more easily accommodates Higgs data \cite{LHCHiggsCrossSectionWorkingGroup:2016ypw,Dawson:2018dcd,Bechtle:2020pkv}, the latter is more appealing as it dynamically motivates a light scalar and addresses naturalness \cite{Csaki:2015hcd,Cacciapaglia:2020kgq}.

\begin{figure}
  \centering
  \includegraphics[width=0.55\textwidth]{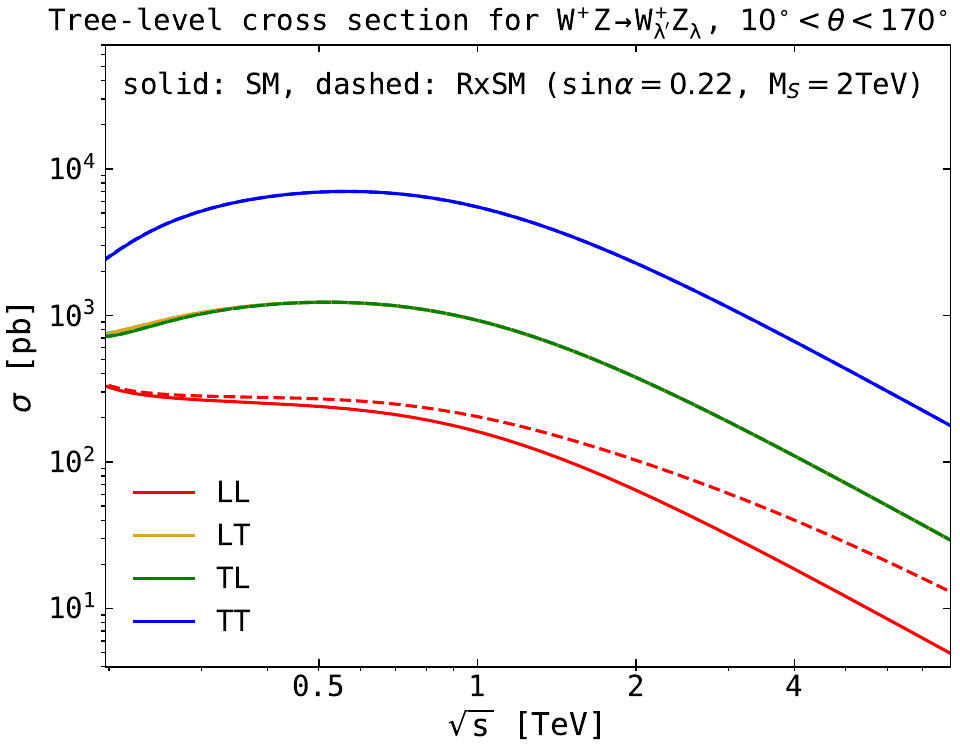}
  \caption{ Tree-level total cross sections in the SM and in its real-singlet extension (RxSM)
    for the on-shell $\PW\PZ\rightarrow\PW_{\lambda}\PZ_{\lambda'}$ scattering, as functions of the
    di-boson CM energy.
    The two initial bosons are unpolarised, while the final ones are either longitudinal ($\rL$)
    or transverse ($\rT$). Numerical results have been obtained with \recola2 \cite{Denner:2017wsf} for the SM
    and for the $\mc Z_2$-symmetric RxSM with $\sin\alpha=0.22,\,\MH
    =125\GeV$, and $M_{\rm S}=2\TeV$. Note that the curves for TL and
    LT are on top of each other.
  }\label{fig:2to2}
\end{figure}

In the case of additional heavy scalars mixing with the SM 125\GeV
Higgs boson, the scale of perturbative-unitarity restoration is
shifted to energies higher than $\Lambda_{\rm ew}$.
To give an example, we consider the $\mc Z_2$-symmetric real Higgs-singlet extension (RxSM) of the SM \cite{Schabinger:2005ei,Bowen:2007ia,Pruna:2013bma,Robens:2016xkb}, characterised by an additional scalar $S$ interacting with the SM light Higgs and EW gauge bosons.
In \reffi{fig:2to2} we show the energy dependence of the $\PW^+\PZ\rightarrow\PW^+\PZ$ cross section
in the SM and in a specific realisation of the RxSM model for
final on-shell bosons with definite polarisation states and
initial unpolarised ones. We consider the academic scenario with a
heavy-scalar mass $M_S=2\TeV$ and a mixing angle $\sin\alpha=0.22$.
It is striking how the transverse--transverse and mixed states are hardly affected by the
dynamics of the scalar sector, with tiny differences just at low energies.
The production of two longitudinal bosons is instead heavily impacted by the presence of
the mixing of the SM Higgs boson with the heavy scalar $S$. Indeed, the fall-off of
the cross section is slower compared to the SM scenario, highlighting the unitarisation of the
model at higher energies than $\Lambda_{\rm ew}$.
This example makes it transparent why the scattering of longitudinal bosons represents such a
crucial probe of the EWSB mechanism realised in nature. 

Nevertheless, EW bosons are unstable particles and experiments can just access them indirectly via
their decay products. Only the full $2\rightarrow 6$ process in \refeq{eq:procdef} can actually be measured,
where both resonant and non-resonant diagrams contribute: The $2\rightarrow 2$ scattering described in this section
is just a sub-process in a sub-set of resonant diagrams that constitute the VBS topologies.
Luckily, the latter have a very characteristic kinematic signature that helps isolating them from
the background topologies contributing to the full  $2\rightarrow 6$ process.
Indeed, VBS is characterised by two jets that are well separated in rapidity and form a system with a large invariant mass.
The two bosons ($\PW\PZ$ for the case at hand) are produced rather centrally and with a quite large
average di-boson invariant mass ($M_{\PW\PZ} \approx 400\GeV$ \cite{Denner:2019tmn}). 
The VBS contributions are enhanced with respect to other contributions by two $t$-channel massive
vector-boson propagators. Moreover, the $t$-channel diagrams in
the $2\rightarrow2$ sub-process provide an additional enhancement
\cite{Denner:1997kq}.
This dominant kinematic configuration leads to sizeable EW radiative corrections coming from large Sudakov logarithms generated by soft and collinear EW-boson exchanges in loop corrections \cite{Denner:2000jv,Denner:2001gw}.
Specifically for $\PW\PZ$ scattering at the LHC, it has been shown
\cite{Denner:2019tmn} that the leading-logarithmic approximation of EW
one-loop corrections captures very accurately the exact NLO EW
corrections to the full off-shell cross section, amounting to $\approx-17\%$ in a typical fiducial LHC setup \cite{CMS:2019uys}. 

In spite of a diminished sensitivity, studying polarised-boson signals in VBS at the LHC remains
a golden channel to get (indirect) access to the EWSB with experimental data, motivating
the recent analyses with Run-2 LHC data \cite{CMS:2020etf,ATLAS:2025wuw}.
Such experimental investigations rely on polarised-template fits, namely
fits of data with several VBS-signal templates, one for each polarisation state
of the produced EW bosons. This required quite an effort in the theoretical community
to properly define polarised templates in currently used simulation tools.  
Specifically for VBS, the modelling beyond leading order
\cite{Ballestrero:2017bxn,Ballestrero:2019qoy,BuarqueFranzosi:2019boy,Ballestrero:2020qgv}
has been achieved only recently \cite{Denner:2024tlu}.
In the next section we are sketching the main steps required for a sound
definition of polarised signals for intermediate EW bosons in LHC processes.

\subsection{Signal definition for polarised vector bosons}\label{sec:signaldef}
A standard for the definition of polarised-boson signals has been
well established in the last years
\cite{Ballestrero:2017bxn,Denner:2020bcz,Poncelet:2021jmj,
  Denner:2021csi,Le:2022lrp,Pelliccioli:2023zpd,Denner:2023ehn,
  Dao:2023kwc,Denner:2024tlu}. It relies on the pole approximation (PA)
\cite{Aeppli:1993cb,Aeppli:1993rs,Denner:2000bj,Denner:2019vbn}
and the separation of intermediate-boson helicity states at the amplitude level:
The former allows to isolate resonant contributions in the EW bosons from the full off-shell process
in a gauge invariant way, which also retains partial off-shell effects, while
the latter is used to extract the individual polarised-amplitude contributions. 
This is the strategy applied in this work.
For illustrative purposes, we sketch in the following the main steps to achieve a polarised-signal definition
for a single EW boson, but the whole discussion applies to the production
and decay of an arbitrary number of EW bosons.

First, as part of the PA, only Feynman diagrams with an $s$-channel $V$-boson propagator
with momentum $k$ connecting a production sub-amplitude $\mc P$ and a
decay sub-amplitude $\mc D$ are selected: Their sum defines 
the {\it resonant} unpolarised amplitude $\cM_{\rm res}$,
which at this level still violates gauge invariance.

In a second step, the PA provides a prescription to define a gauge-invariant {\it unpolarised amplitude}
$\cM_{\rU}$ starting from $\cM_{\rm res}$ as
\begin{align}
\cM_{\rm res}\,&=\,
\mc P_{\mu}(k)\,
\frac{-g^{\mu\nu}}{k^2-M_V^2+\ri M_V \Gamma_V}\,
\mc D_{\nu}(k)
\,\longrightarrow\,
\mc P_{\mu}(\tilde{k})\,
\frac{-g^{\mu\nu}}{k^2-M_V^2+\ri M_V \Gamma_V}\,\mc D_{\nu}(\tilde{k})\,\equiv \cM_{\rU}\,,
\label{eq:amp_unp}
\end{align}
where the off-shell boson momentum $k$ ($k^2\neq M_V^2$) has been
replaced in the amplitude numerator by the on-shell-projected momentum $\tilde{k}$
($\tilde{k}^2= M_V^2$), while the off-shell factor in the denominator is kept with
the original kinematics.
The on-shell mapping $\{k\rightarrow \tilde{k}\}$ is not unique \cite{Denner:2002cg,Denner:2024xul}
and depends on the kinematic quantities that are preserved. In our calculation we apply
the same prescription as the one of \citere{Denner:2024tlu}.

Third, once the intermediate EW bosons have been set on shell, only the physical polarisation states
$\lambda$ can contribute, namely the longitudinal, left-handed and
right-handed modes.%
\footnote{The contribution from the unphysical scalar polarisation of the vector boson
  cancels exactly with the Goldstone-boson one for on-shell
  amplitudes. Both even vanish in the case of massless final-state fermions.}
A {\it polarised amplitude} $\cM_{\lambda}$ for each polarisation state can
be extracted from the individual terms of the sum over all physical polarisation states entering $\cM_{\rU}$:
\begin{align}
\cM_{\rU}\,&=\,
\mc P_{\mu}(\tilde{k})\,
\frac{-g^{\mu\nu}}{k^2-M_V^2+\ri M_V \Gamma_V}\,\mc D_{\nu}(\tilde{k})
\equiv\,
\mc P_{\mu}(\tilde{k})\,
\frac{ 
\sum_{\lambda'}\varepsilon^{\mu}_{\lambda'}(\tilde{k})\varepsilon^{\nu\,*}_{\lambda'}(\tilde{k})}{k^2-M_V^2+\ri M_V \Gamma_V}\,\mc D_{\nu}(\tilde{k})\nnb\\
&\longrightarrow\,
\mc P_{\mu}(\tilde{k})\,
\frac{ 
\varepsilon^{\mu}_\lambda(\tilde{k})\varepsilon^{\nu\,*}_{\lambda}(\tilde{k})}{k^2-M_V^2+\ri M_V \Gamma_V}\,\mc D_{\nu}(\tilde{k})\,\equiv \cM_{\lambda}\,.
\label{eq:amp_split}
\end{align}

Fourth and last, by taking the modulus squared of the polarised amplitude $\cM_{\lambda}$ defined
in \refeq{eq:amp_split}, the proper polarised matrix-element weight is obtained. The latter
represents the main ingredient
to compute generic (integrated or differential) cross sections, upon multiplication of
suitable factors (symmetry, flux) and a phase-space weight. It is worth mentioning that
the phase-space weights in the PA are evaluated with the original off-shell kinematics.

An interesting aspect is that, if the modulus squared is applied at the level of the unpolarised amplitude $\cM_{\rU}$,
we do not only get an incoherent sum
of squared polarised amplitudes, but also a number of off-diagonal terms coming from the {\it interference\/} between different
modes:
\beq
|\cM_{\rU}|^2\,=\,\sum_{\lambda}|\cM_{\lambda}|^2
+ \sum_{\lambda \ne \lambda'}\cM_{\lambda}^* \cM_{\lambda'}\,.
\eeq
While the unpolarised amplitude is as a whole Lorentz covariant,
the polarised amplitudes are not. Therefore,
the polarised-boson signals have to be defined in a specific reference frame.
In this article we work in the so-called {\it modified helicity coordinate system} \cite{Aaboud:2019gxl}, which understands polarisation states defined in the di-boson CM frame.
This is a natural choice for VBS processes, as it represents the reference frame where
the $\PW\PZ \rightarrow \PW\PZ$ sub-process takes place (the two bosons are back to back).

In this work we consider longitudinal ($\rL$) and transverse ($\rT$) modes of $\PW$ and $\PZ$ bosons.
While a longitudinal-boson signal is obtained by selecting one single
term in the vector-boson-propagator numerator, the transverse mode is defined as the coherent
sum of the left- and right-handed polarisation modes of a vector boson, including the
interference between the two~\cite{Ballestrero:2019qoy,Denner:2020bcz}.

The general approach described so far applies not only at LO but also when adding
radiative corrections of QCD or EW nature. Although conceptually straightforward, the extension
to higher orders in perturbation theory requires a number of technical precautions when
dealing with intermediate polarised EW vector bosons.
A first crucial aspect is that all matrix elements entering a next-to-leading- (or higher-) order
calculation, of both virtual and real origin, have to be evaluated in the same reference frame \cite{Denner:2020bcz}.
A second important requirement is that the PA procedure must not disrupt the subtraction of infrared (IR) singularities in the local and integrated counterterms \cite{Denner:2021csi}.
This must be achieved with a careful selection of singular regions associated to the production and to the decay of EW bosons
and with a proper application of the on-shell mapping to the Catani--Seymour counterterms entering the subtracted-real contribution \cite{Denner:2021csi}.
Furthermore, the case of photon radiation off intermediate $\PW$ bosons has to be treated
including additional subtraction counterterms \cite{Le:2022ppa,Denner:2023ehn,Denner:2024tlu}: Radiation from
charged resonances can indeed contribute both to the production and the decay of the EW bosons.
For a broad and detailed discussion of all these technical aspects we refer to \citere{Denner:2024tlu}.

\subsection{Treatment of triply resonant contributions}\label{sec:tribosons}
The VBS processes overlap with triple-boson production, which is typically suppressed
by the selection cuts used to isolate vector-boson-fusion event topologies.
This overlap is particularly problematic when applying the DPA \cite{Denner:2024xul,Denner:2024tlu}.
Our study targets the polarisation structure of two
leptonically-decaying EW bosons, one $\PW$ and one $\PZ$, which therefore have to be projected on shell.
In order to preserve gauge invariance,
the DPA approach requires to set to zero the EW-boson widths everywhere in the matrix element but
in the propagator factors of the two target bosons. This operation leads to divergences
in partonic contributions that allow for an additional $s$-channel EW boson,
whose propagator pole is not protected anymore by a finite width.

At the $\mathcal{O}(\alpha^6)$, the VBS selection cut on the invariant mass
of the two hardest jets completely removes any divergent contribution,
since $M_{\Pj_1\Pj_2}>M_{\rm cut}\gg\MZ$ (see \refse{sec:ATLASfid}).
The same cut also suppresses most of the potentially singular contributions at NLO accuracy,
but still does not prevent some rare events from spoiling the integration.
This occurs for instance in real-radiation events with a third hadronically decaying
EW boson characterised by a virtuality that is very close to the pole mass, and a hard photon
that gets recombined with one of the two quarks from the boson decay itself.
The resulting invariant mass of the tagging-jet system, \ie $M_{\Pj_1\Pj_2}$,
might exceed the $M_{\rm cut}$ threshold, in which case the divergent event would pass
the selection cuts.
Even though Feynman diagrams containing a third EW $s$-channel resonance are typically
associated with small contributions to the total cross section, 
still the presence of large-weight events can spoil the convergence of the result,
especially at the differential level.
We stress that this issue would be even more severe if QCD or photon-induced corrections
were included at NLO, owing to the presence of three QCD partons in the final state, potentially
leading to analogous unprotected propagator poles
\cite{Denner:2024tlu}.
Therefore, a proper handling of triply resonant terms is required.

In \citere{Denner:2024tlu} the unprotected $s$-channel resonances were
regularised by restoring the finite EW-boson width in the DPA matrix
elements of the real and local subtraction terms.  This approach,
although simple, has the drawback of introducing a gauge dependence in
the result, whose numerical impact has to be verified case by case.
Moreover, for diagrams where a resonance treated in the PA emits
photons, which is relevant here, the internal resonance coupling to
these photons gets a width as well. As a consequence the corresponding
soft singularities are regularised by the width. Since these singularities
in our calculation are already accounted for by dedicated subtraction
terms based on dimensional regularisation \cite{Denner:2024tlu}, this spoils the IR
subtraction.

Therefore, we propose here a different regularisation approach inspired by
\citeres{Baur:1991pp,Kurihara:1994fz,Argyres:1995ym} 
and based on the usage of a \emph{fudge factor} $\mathcal{F}$,
  defined as
\begin{align}\label{eq:fudge_factor}
\mathcal{F}(s_V)=
\begin{cases}
  1 & |\sqrt{s_V}-M_V|>c_{\mc F}\,\Gamma_V\\
  \frac{(s_V-M^2_V)^2}{(s_V-M^2_V)^2+\Gamma^2_VM^2_V} &  |\sqrt{s_V}-M_V|<c_{\mc F}\,\Gamma_V\\
  \end{cases},
\end{align}
where  $s_V$ refers to the Lorentz invariant associated with the unprotected resonance $V$
($=\PW,\,\PZ$) decaying to a quark--antiquark pair, and $c_{\mc F}$ is some parameter.
For partonic channels allowing for an additional $s$-channel resonance $V$, the real matrix element
gets multiplied by the factor $\mc F$.
Close to the pole mass ($s_V\rightarrow M^2_V$), the fudge factor effectively
replaces the unregularised propagator with the one including a finite width $\Gamma_V$.
Since the complete gauge-invariant amplitude is multiplied by the
fudge factor, this does not introduce any gauge dependence.
In order not to spoil the cancellation of IR singularities,
the same fudge factor must be applied to the local subtraction counterterms.
Moreover, to avoid any mismatch between local and integrated counterterms, the same
factor is applied to integrated counterterms as well.
Note that the fudge factor switches off the contributions that do not
include the corresponding resonance, \ie it introduces an error of at
least $\Gamma_V/M_V$. In our case this is a very small
effect, as it only concerns contributions at NLO in restricted regions
of phase space.

Although $\mathcal{F}\to{}1$ by construction for $s_V\gg M^2_V$, the range of
application of the fudge factor can be restricted to $s_V$ values that
are less than $c_{\mc F}$ widths away from the resonance pole
mass.  We have numerically checked that by varying the parameter
$c_{\mc F}$ (between 3 and infinity) the result does not change
within integration errors. Nonetheless, applying the fudge factor only in the
vicinity of the resonance pole mitigates the risk of miscancellations
between the real contributions and the corresponding local-subtraction
counterterms, for which the fudge factors are constructed using the
real and the mapped Born kinematics, respectively.  The definition of
$\mathcal{F}$ in \refeq{eq:fudge_factor} ameliorates this problem to a
large extent.

\subsection{Input parameters}\label{sec:setup}
The Run-3 energy of the LHC is considered, \ie $\sqrt{s} =
13.6\TeV$. The on-shell values for masses and widths of EW bosons
are taken from the latest PDG review \cite{ParticleDataGroup:2024cfk},
\begin{align}\label{eq:EWmasses}
  \MZOS    &{}=  91.1880\GeV,\quad & 
  \GZOS    &{}=  2.4955  \GeV,\nonumber \\ 
  \MWOS    &{}=  80.3692 \GeV,\quad & 
  \GWOS    &{}=  2.085  \GeV\,, 
\end{align}
and are converted into pole values ($\MZ,\MW,\GW,\GZ$) following \citere{Bardin:1988xt}.
The parameters used for the top quark and the Higgs boson read \cite{ParticleDataGroup:2024cfk}
\begin{align}
\Mt &{}= 172.57\GeV,\quad & 
\Gt &{}= 1.42\GeV, \nnb\\  
\MH &{}= 125.20\GeV,\quad & 
\GH &{}= 0.0041\GeV\,.  
\end{align}
The Higgs width is taken from \citere{LHCHiggsCrossSectionWorkingGroup:2016ypw}.
The $G_\mu$ input scheme \cite{Denner:2000bj,Dittmaier:2001ay} is used to calculate the EW coupling.
The off-shell calculation is carried out employing the complex-mass scheme
\cite{Denner:2005fg,Denner:2006ic},
\begin{equation}\label{eq:alphaOFFSH}
  \alpha =
  \frac{\sqrt{2}}{\pi}\,G_\mu
  \,\left|\mu_{\PW}^2\,
  \left(  1 - \frac{\mu_{\PW}^2}{\mu_{\PZ}^2} \right)
  \right|\,,
  \qquad \mu_V^2=M_V^2-\ri M_V\Gamma_V\,, \quad V=\PW,\,\PZ\,.
\end{equation}
Consistently with the DPA prescription, the (un)polarised results are obtained using real couplings, leading to
\begin{equation}
  \alpha =
  \frac{\sqrt{2}}{\pi}\,G_\mu\,\MW^2\,
  \left(
  1 - \frac{\MW^2}{\MZ^2}
  \right).
\end{equation}
For both off-shell and DPA calculations, the Fermi constant is set to 
\begin{align}
\GF = 1.1663788 \times 10^{-5}\GeV^{-2}\,.
\end{align}

The \sloppy\texttt{NNPDF40\_nnlo\_as\_01180\_qed}~\cite{NNPDF:2024djq} PDF set is used.
The renormalisation and factorisation scales are defined dynamically,
\begin{equation}
\label{eq:scale}
\mu_{\rm R} =
\mu_{\rm F} =
\sqrt{\,\pt{\Pj_1}\,\pt{\Pj_2}} \,,
\end{equation}
where $\Pj_1$ and $\Pj_2$ are the two leading jets according to transverse-momentum ordering.

\subsection{Event selection}\label{sec:ATLASfid}
While considering the Run-3 LHC collision energy ($\sqrt{s} = 13.6\TeV$), we apply selection cuts inspired by the ATLAS fiducial volume used for the Run-2 data analysis of \citere{ATLAS:2024ini}.
We require three charged leptons and missing transverse momentum (defined as the truth-neutrino $p_{\rm T}$), fulfilling
\beqn\label{eq:fidLEP}
&&
\pt{\Pe^+}>20\GeV\,,\qquad \pt{\mu^\pm}>15\GeV\,,\qquad
|\eta_{\Pe^+}|<2.5\,,\qquad |\eta_{\mu^\pm}|<2.5\,,\nnb\\
&& M_{\rT,\PW}>30\GeV\,,\qquad 
|M_{\mu^+\mu^-}-\MZOS|< 10\GeV\,,\nnb\\
&&\Delta R_{\Pe^+\mu^\pm}>0.3\,,\qquad \Delta R_{\mu^+\mu^-}>0.2\,,
\eeqn
where the transverse mass is defined as $M_{\rT,\PW} = \sqrt{2\,\pt{\Pe^+}\pt{\nu_{\Pe}}(1-\cos\Delta\phi_{\Pe^+\nu_{\Pe}})}$.
We also ask for at least two jets such that
\beqn\label{eq:fidJET}
&&
\pt{\Pj}>40\GeV\,,\qquad |\eta_{\Pj}|<4.5\,, \qquad \Delta R_{\Pj\ell}>0.3\quad(\ell=\Pe^+,\mu^\pm)\,.
\eeqn
The two hardest jets passing the selections in~\refeq{eq:fidJET} and satisfying $\eta_{\Pj_1}\cdot\eta_{\Pj_2}<0$
need to fulfil
\beqn
&& M_{\Pj_1\Pj_2}>500\GeV\,.
\eeqn
All recombinations are performed for particles with $|\eta|<5$.
First, photons are recombined with charged leptons with resolution
radius $R=0.1$ using the Cambridge--Aachen algorithm \cite{Dokshitzer:1997in}.
Then, QCD partons and possible photons that do not dress the leptons are clustered into jets with resolution radius $R=0.4$ using the anti-$k_{\rm t}$ algorithm \cite{Cacciari:2008gp}.

\section{Results}\label{sec:results}
In this section we collect the numerical results obtained with the
\mocanlo multi-channel-integration code, which has already been
employed for several NLO calculations with polarised vector bosons
\cite{Denner:2020bcz,Denner:2020eck,Denner:2021csi,Denner:2022riz,Denner:2023ehn,Denner:2024tlu,Carrivale:2025mjy}.
All results (full off-shell, DPA polarised and unpolarised) have been reproduced at both
integrated and differential level by an independent calculation with the \bbmc code,
which has also been used for a number of polarised-boson calculations \cite{Denner:2022riz,Denner:2023ehn,Denner:2024tlu,Carrivale:2025mjy}.
Both codes rely on tree-level and one-loop SM amplitudes provided by the \recola~1 library \cite{Actis:2012qn,Actis:2016mpe} with loop-integral reduction and calculation carried out through \collier \cite{Denner:2016kdg}. The two codes embed independent implementations of the dipole formalism for the subtraction of IR singularities \cite{Catani:1996vz,Dittmaier:1999mb,Catani:2002hc,Dittmaier:2008md}.

In the following we only present results obtained for the Run-3 LHC
collision energy (13.6\TeV). However, we have checked that
the analogous results at the Run-2 energy (13\TeV) do not show any
substantial difference, up to the obvious overall change in
normalisation of the total cross sections.  In \refses{sec:integrated}
and \ref{sec:differ} we focus on integrated and differential
results, respectively.

\subsection{Integrated cross sections}\label{sec:integrated}
We present the integrated cross sections of the VBS signals
depending on the polarisation of the two resonant EW
vector bosons in \refta{tab:obs_1}, showing results at LO,
i.e $\order{\alpha^6}$, and at NLO EW, i.e $\order{\alpha^7}$.
\begin{table}
  \begin{center}
    \begin{tabular}{c|ccc|cc}
      \hline\rule{0ex}{2.7ex}
      \cellcolor{blue!9} mode & \cellcolor{blue!9} $\sigma_{\rm LO}$ [fb] & \cellcolor{blue!9} $\delta_{\rm EW}$ [\%]  & \cellcolor{blue!9} $\sigma_{\rm NLO\,EW}$ [fb] & \cellcolor{blue!9}$f_{\rm LO}$ [\%] &\cellcolor{blue!9}  $f_{\rm NLO\,EW}$ [\%] \\
      \hline\rule{0ex}{2.7ex}
      full     & $0.27208(1)$  & $-18.5$ & $0.2216(2)$     & $101.6$ & $101.5$ \\
      unp.     & $0.26777(1)$  & $-18.5$ & $0.21832(3)$    & $100.0$ & $100.0$ \\
      $\rL\rL$ & $0.023045(2)$ & $-14.5$ & $0.019701(4)$   & $  8.6$ & $  9.0$ \\
      $\rL\rT$ & $0.049365(3)$ & $-17.9$ & $0.040516(8)$   & $ 18.4$ & $ 18.6$ \\
      $\rT\rL$ & $0.049023(3)$ & $-17.9$ & $0.040233(8)$   & $ 18.3$ & $ 18.4$ \\
      $\rT\rT$ & $0.141725(8)$ & $-19.6$ & $0.11395(2)$    & $ 52.9$ & $ 52.2$ \\
      int.     & $0.00462(2)$  & $-15.0$ & $0.00393(4)$    & $  1.7$ & $  1.8$ \\
      \hline
    \end{tabular}\qquad
  \end{center}
  \caption{Integrated cross sections for $\PW^+\PZ$ scattering at the
    LHC at $13.6\TeV$ in the fiducial setup described
    in~\refse{sec:ATLASfid}. The LO cross section, the NLO\,EW
    corrections relative to LO, and the NLO EW cross section for the
    various unpolarised and polarised signals are presented in the
    2nd, 3rd, and 4th column, respectively. The first/second
    polarisation label refers to the $\PW^+\!/\PZ$ boson. The last row shows the interference contribution. The two rightmost columns list the polarisation fractions at LO and at NLO EW, defined as ratios of the corresponding cross sections over the unpolarised one.\label{tab:obs_1}
  }
\end{table}
The NLO-accurate cross sections are defined according to:
\beq
\sigma_{\rm{NLO\,EW}} = \sigma_{\rm LO} \left(1 + \delta_{\rm EW} \right)\,.
\eeq
In the two rightmost columns the LO and NLO EW joint polarisation fractions are shown.
These are defined as ratios of doubly polarised cross sections over the unpolarised one:
\beq\label{eq:polfrac}
f_{k,\lambda\lambda'} = \frac{\sigma_{k,\lambda\lambda'}}{\sigma_{k,\rm unp.}}\,,\qquad \lambda,\lambda'=\rL,\rT,\qquad k ={\rm LO,\,NLO\,EW}\,.
\eeq
Together with polarised signals, we show integrated results for
the unpolarised signal in the DPA (unp.) and for
the full off-shell description of the process  (full).
The difference between the full off-shell and the unpolarised DPA description provides an estimate of
the genuine off-shell effects that are not captured by the DPA,
including non-resonant diagram topologies
and terms beyond the leading one in the pole expansion \cite{Denner:2019vbn}.
Note that the polarisation fractions in \refeq{eq:polfrac} are normalised
to the unpolarised DPA cross section, not to the full off-shell one.
The interference contribution (int.) given in the last row of \refta{tab:obs_1} is obtained 
by subtracting the sum of polarised cross sections from the unpolarised DPA one, namely,
\beq\label{eq:int}
\sigma_{k,\rm int.} = \sigma_{k,\rm unp.}-
\sum_{\lambda,\lambda'=\rL,\rT}\sigma_{k,\lambda\lambda'}\,,
\qquad k ={\rm LO,\,NLO\,EW}\,.
\eeq

The off-shell effects amount to $1.6\%$ at LO and $1.5\%$ at NLO EW,
which is well inside the expected range of accuracy of the DPA,
namely $\order{\Gamma_\PW/M_\PW}\approx \order{\Gamma_\PZ/M_\PZ}
\approx 3\%$ \cite{Denner:2000bj,Denner:2019vbn}.
In both the full off-shell and the unpolarised DPA calculations the one-loop EW corrections amount to
$-19\%$, of the same size as those found in previous off-shell calculations at an LHC energy of $13\TeV$
\cite{Denner:2019tmn}.
Such large and negative corrections were found for all VBS channels \cite{Biedermann:2016yds,Denner:2019tmn,Denner:2020zit,Denner:2022pwc,Denner:2024tlu} and are known to originate from logarithmic enhancements in soft and collinear configurations of EW bosons in virtual corrections \cite{Denner:2000jv,Denner:2001gw}. 

The dominant logarithmic corrections depend on the polarisation mode of the EW bosons \cite{Denner:2000jv} which leads to relative NLO EW corrections that are different for the doubly polarised signals we consider. The NLO corrections are maximal (about $-20\%$) for the purely transverse mode (TT), they amount to $-18\%$ for mixed states (LT, TL) and $-15\%$ for the purely longitudinal mode (LL). A very similar hierarchy was found in the case of polarised $\PW^+\PW^+$ scattering \cite{Denner:2024tlu}.
This stems from the universal character of such logarithmic EW corrections,
which are known to be smaller for longitudinal bosons than for transverse ones, owing to the difference in the EW Casimir operators multiplying the leading soft-collinear Sudakov double logarithms \cite{Denner:2000jv}.

Looking at the LO polarised cross sections, the purely transverse mode gives the dominant contribution to the unpolarised cross section ($53\%$), the mixed modes give both a $18\%$ contribution, and the purely longitudinal mode is suppressed ($9\%$).
These results are very similar to those found in previous literature \cite{Ballestrero:2019qoy,Ballestrero:2020qgv}. Furthermore, the differences in the EW corrections
do not affect sizeably the polarisation fractions, which remain almost unchanged between LO and NLO EW.
It is worth recalling that the purely longitudinal $\PW\PZ\rightarrow\PW\PZ$ 
process undergoes in the SM delicate cancellations between large, unitarity-violating contributions coming from pure-gauge and from Higgs-mediated diagrams \cite{Veltman:1989vw,Dawson:1989up,Passarino:1990hk,Denner:1997kq}. 
Compared to polarised $\PW^+\PW^+$ scattering, the mixed TL and LT~modes are more prominent in case of $\PW^+\PZ$ scattering, but the general picture with the TT~modes contributing most to the cross section remains identical \cite{Denner:2024tlu}.

The interference term defined in \refeq{eq:int} contributes at the
$2\%$ level. Thus, just like off-shell effects, the interference
impact is well below the expected Run-3 experimental sensitivity in
rare processes like $\PW\PZ$ scattering. However, in light of a
precision study for the upcoming High-Luminosity stage of the LHC,
both off-shell and interference effects should be properly
incorporated in data analyses, \eg as separate templates.

\subsection{Differential cross sections}\label{sec:differ}
In order to appreciate better the results discussed for integrated cross sections,
it is essential to analyse differential observables. This is the topic of this subsection.

\begin{figure}
  \centering
  \subfigure[\label{fig:M4L}]{\includegraphics[width=0.48\textwidth, page=1]{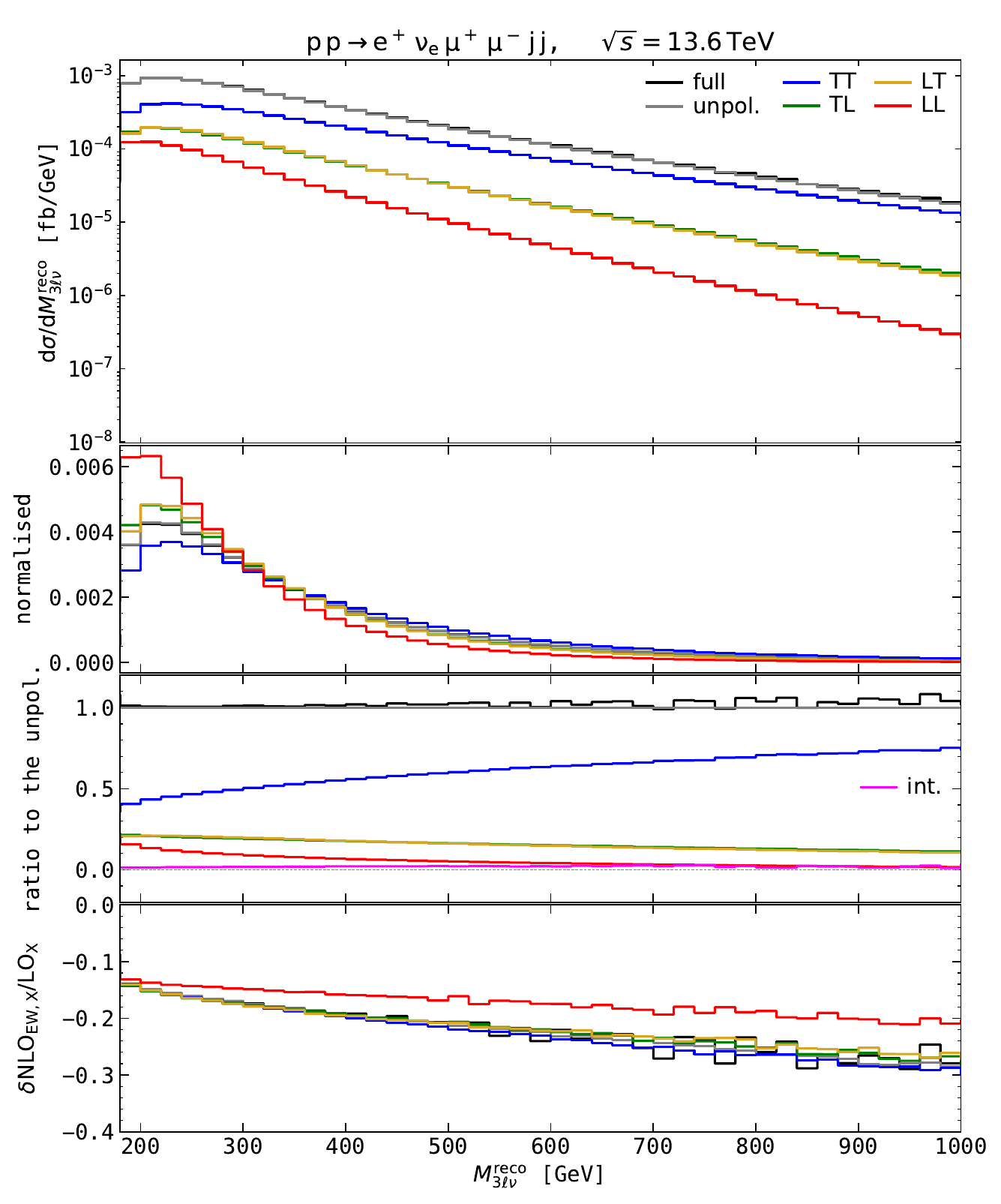}}
  \subfigure[\label{fig:MTWZ}]{\includegraphics[width=0.48\textwidth, page=3]{fig/NLO-output_VBS_combined-PAPER.pdf}}\\
  \caption{
    Distributions in the reconstructed invariant mass (left) and the transverse mass (right)
    of the $\PW\PZ$ system
    for $\PW^+\PZ$ scattering at the LHC in the fiducial setup described in~\refse{sec:ATLASfid}.
    The invariant mass is obtained through the neutrino-reconstruction algorithm
    of \citere{Aaboud:2019gxl}, while the transverse mass is defined in \refeq{eq:MTWZ}.
    The structure reads (from the top down):
    absolute NLO EW distributions,
    normalised NLO EW distributions (to have unit integral),
    ratios over the unpolarised DPA results (at NLO EW accuracy),
    NLO EW corrections relative to LO.
    The colour key is as follows:
    full off-shell (black), unpolarised (grey), LL (red), LT (yellow),
    TL (green), TT (blue), and interference (magenta).
    The first/second polarisation label refers to the $\PW^+\!/\PZ$ boson.}
  \label{fig:first}
\end{figure}

We start by presenting in \reffi{fig:M4L} the distribution in the invariant mass of the di-boson system,
\ie the invariant mass of the system formed by the three charged leptons and the reconstructed neutrino.
Owing to the presence of a single neutrino, it is possible to reconstruct its momentum
analytically by enforcing the $\Pe^+\nu_{\Pe}$ pair to be on the $\PW$-boson mass shell.
This approach leads to either complex or real solutions of a quadratic equation in the longitudinal component
of the neutrino momentum. Following \citere{Aaboud:2019gxl}, in the former case the real part of the complex
solutions is taken. Moreover, between the two real solutions,
the one that minimises the absolute value of the longitudinal component of the neutrino momentum
is chosen.
The polarised distributions in \reffi{fig:M4L} fall off with different rates in the high-energy limit.
As motivated in \refse{sec:2to2} based on unitarity arguments, in the $2\rightarrow 2$ limit with external on-shell bosons all polarised signals must
decrease at least with the inverse squared power of the di-boson CM energy.
Nonetheless, in LHC events VBS topologies are just a subset of resonant diagrams and the di-boson CM energy is modulated by the PDF weights, resulting in a different fall-off of the four polarisation modes.
The asymptotic behaviour found for the various polarisation configurations is very similar to the one of $\PW^\pm\PW^\pm$ scattering
\cite{Denner:2024tlu}: In the high-energy limit the $\rT\rT$ mode dominates over other modes.
The NLO EW corrections for the $\rL\rL$ state are less sizeable than for other modes, owing to the smaller
EW Casimir factors for longitudinal bosons than for transverse ones multiplying the leading Sudakov logarithms \cite{Denner:2000jv,Denner:2001gw}.
They reach about $-20\%$ at $1\TeV$ for $\rL\rL$, whereas they amount to $-30\%$ for all other modes.  
The interference and off-shell effects are small and almost flat over the whole mass range.

In \reffi{fig:MTWZ} we show the distribution in the transverse mass of the leptonic system, defined as in \citere{ATLAS:2016bkj},
\beq\label{eq:MTWZ}
M_{\rT,\PW\PZ} = \sqrt{\Big(
  \sum_{i=\ell,\nu} |\vec{\,p}_{\rT,i}|\Big)^2 -
  \Big|
  \sum_{i=\ell,\nu} \vec{\,p}_{\rT,i}\Big|^2
}\,.
\eeq
This observable is sensitive to potential beyond-the-SM effects in its high-energy tail,
and has been thoroughly exploited in inclusive $\PW\PZ$ measurements \cite{ATLAS:2016bkj,Aaboud:2019gxl}
as well as in associated production with two jets \cite{CMS:2019uys,ATLAS:2024ini}.
At variance with the di-boson invariant mass [see \reffi{fig:M4L}], the transverse mass in \refeq{eq:MTWZ}
can be fully reconstructed in LHC events and turns out to be a good proxy of $M_{3\ell\nu}$ especially
in the high-energy regime.
While the tails are more severely suppressed than for $M_{3\ell\nu}$, the relative contributions from different polarisation modes
in the high-energy regime closely resemble the findings in the $M_{3\ell\nu}$ distributions.
In the most populated region, namely the $200\GeV$-wide window centered at $M_{\rT,\PW\PZ}\approx \MW+\MZ$, we observe
 rather different shapes for the polarisation modes, with the $\rL\rL$ signal featuring a narrower
peak than the others. Additionally, the distribution maximum for the
$\rL\rL$ mode is at $160\GeV$,
while for the $\rT\rT$ one it is at $180\GeV$. The NLO EW corrections for the $\rL\rL$ signal above $\MW+\MZ$
are rather flat up to $500\GeV$, while for other modes they continuously decrease from $-15\%$ to $-35\%$
in the considered range.
While the off-shell effects are small over the whole distribution, large
interference terms are non negligible. These are positive above
$\MW+\MZ$ and increasingly large and negative below, where all distributions
are strongly suppressed.

\begin{figure}
    \centering
    \subfigure[\label{fig:PTZ}]{\includegraphics[width=0.48\textwidth, page=2]{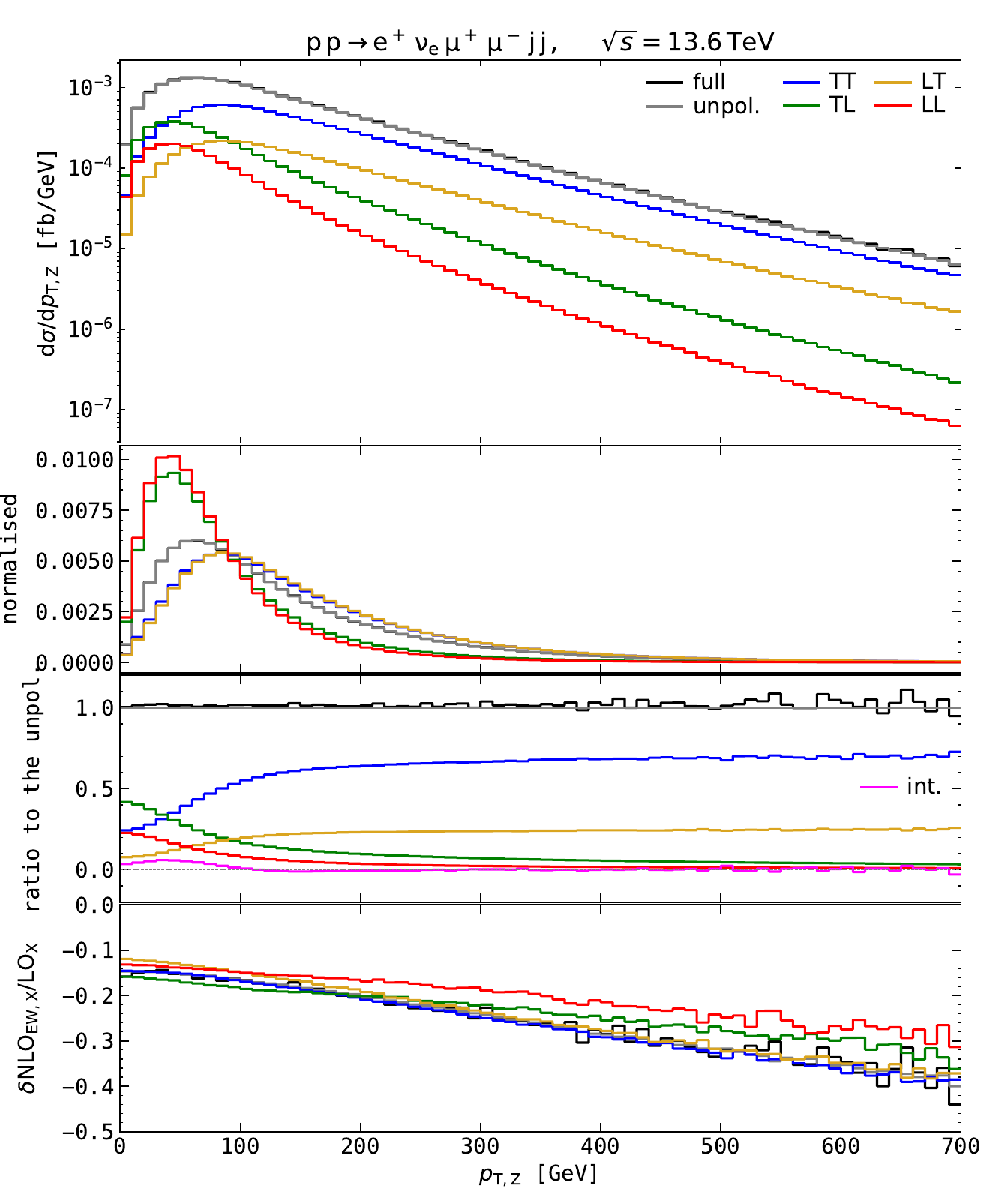}}
    \subfigure[\label{fig:PTJJ}]{\includegraphics[width=0.48\textwidth, page=9]{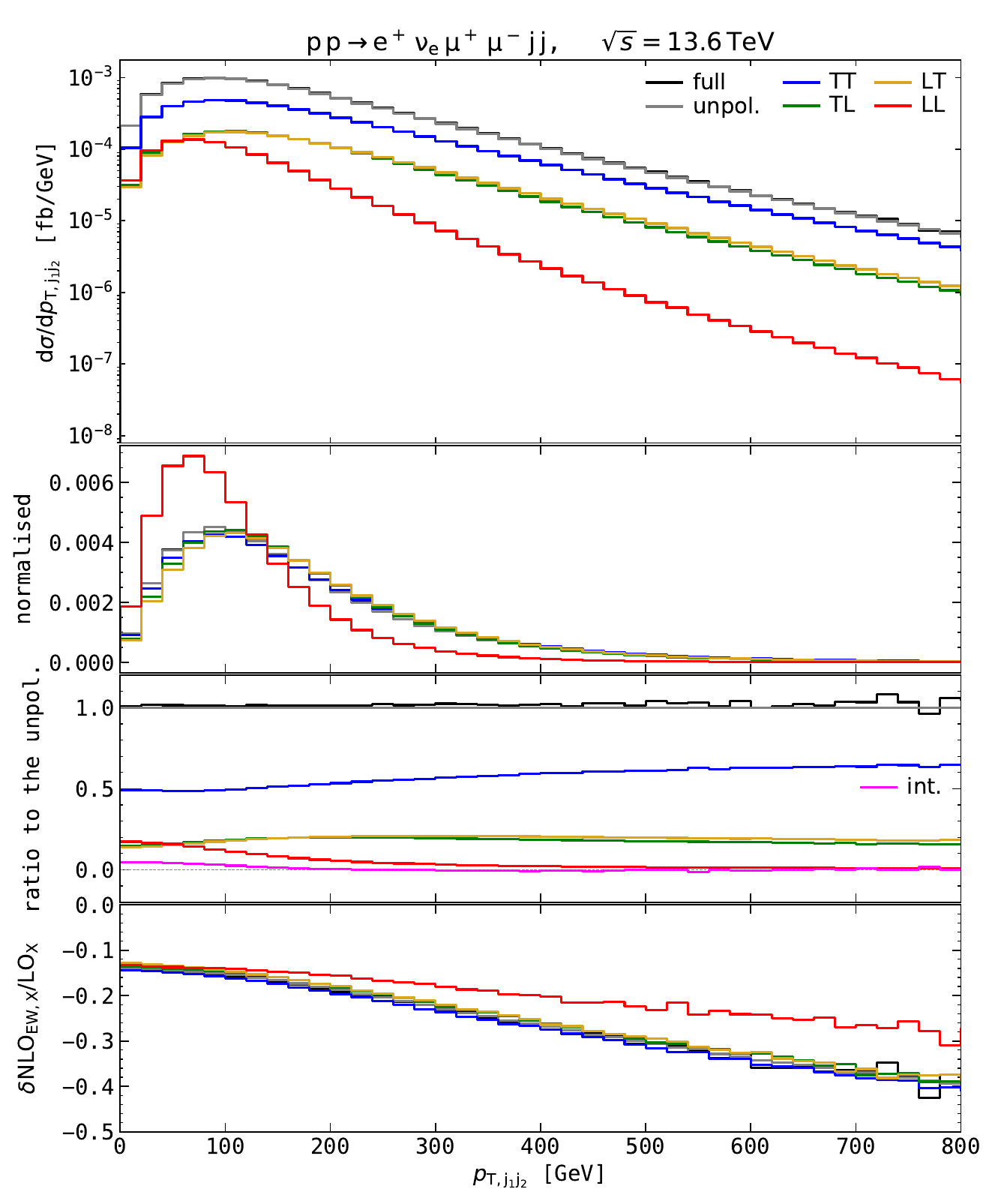}}\\
    \caption{
      Distributions in the transverse momentum of the muon--antimuon pair (left)
      and of the tagging-jet system (right) 
      for $\PW^+\PZ$ scattering at the LHC in the fiducial setup described in~\refse{sec:ATLASfid}.
      Same structure as in \reffi{fig:first}.
    }\label{fig:second}
  \end{figure}

We now turn to transverse-momentum distributions of the $\PZ$ boson, \ie the muon--antimuon system, in \reffi{fig:PTZ},
and of the system of the two tagging jets in \reffi{fig:PTJJ}.
In the former case, the $\rL\rL$ and $\rT\rL$ modes, both involving a longitudinal $\PZ$ boson,
show the same asymptotic behaviour, with a stronger asymptotic decrease 
than the $\rL\rT$ and $\rT\rT$ ones (with a transverse $\PZ$ boson).
The normalised shapes confirm
\cite{Ballestrero:2017bxn,Ballestrero:2019qoy,Ballestrero:2020qgv,Denner:2024tlu}
that a longitudinal boson in VBS tends to be produced
with smaller transverse momentum ($\pt{\PZ}\approx 40\GeV$) than the transverse one ($\pt{\PZ}\approx 90\GeV$). The relative NLO corrections show the typical pattern expected from large EW Sudakov logarithms, representing the leading effects to increasingly large negative corrections.
A very similar behaviour in the NLO EW corrections is also found in the distributions in $\pt{\Pj_1\Pj_2}$, with the $\rL\rL$ mode reaching up to $-30\%$ at about $800\GeV$, and all other modes up to $-40\%$
in the same region, owing to the different EW Casimir operators multiplying the leading logarithms of virtual origin. At variance with $\pt{\PZ}$, which is sensitive to the polarisation of the $\PZ$ boson, the $\pt{\Pj_1\Pj_2}$ observable provides a marked discrimination power between the $\rL\rL$ signal and all other ones: The $\rL\rL$ distribution gets suppressed faster than other polarisation modes, as already observed in $\PW^\pm\PW^\pm$ scattering \cite{Ballestrero:2020qgv}.  

\begin{figure}
  \centering
  \subfigure[\label{fig:CTSTAR}]{\includegraphics[width=0.48\textwidth, page=7]{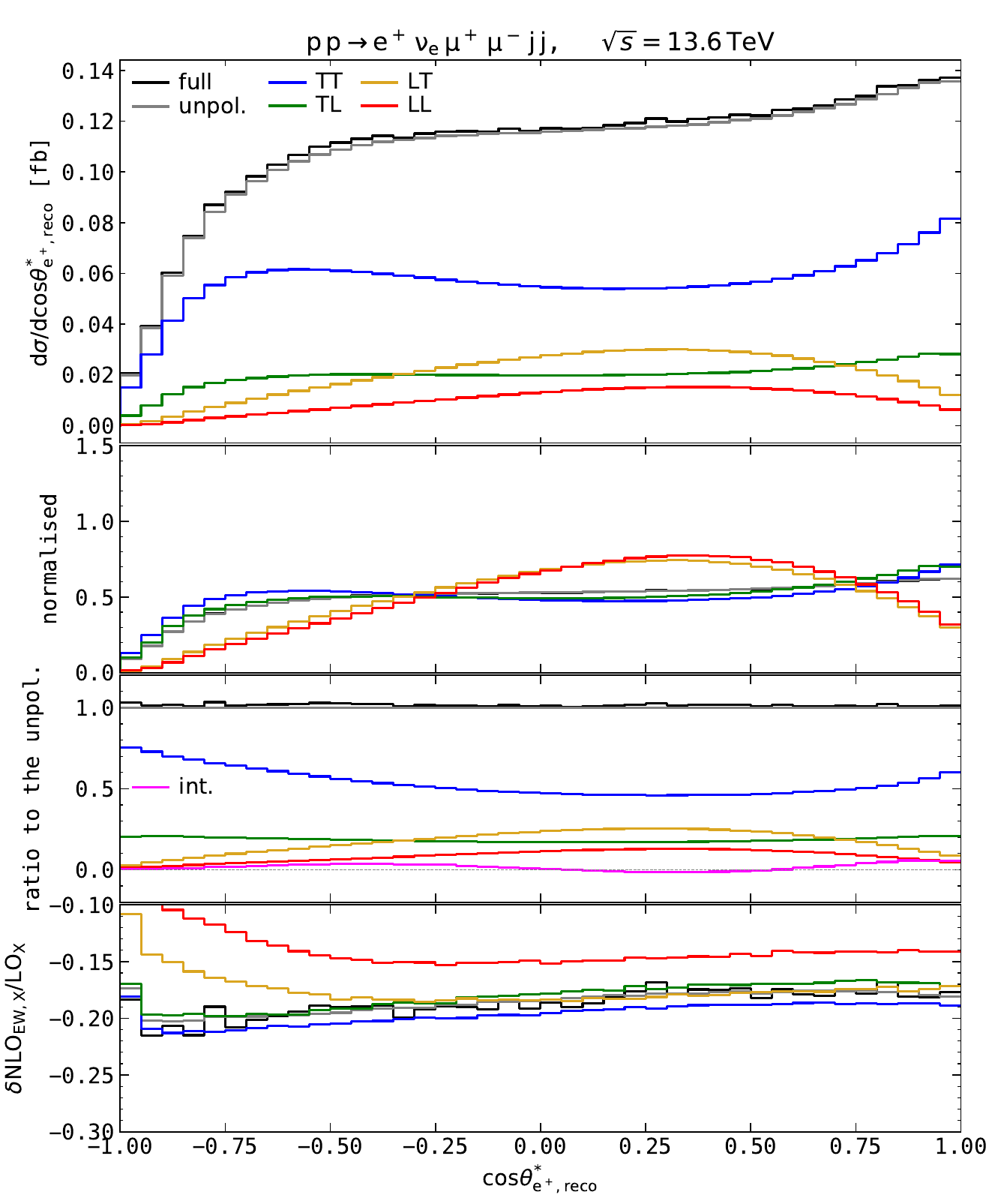}}
  \subfigure[\label{fig:DPHILV}]{\includegraphics[width=0.48\textwidth, page=5]{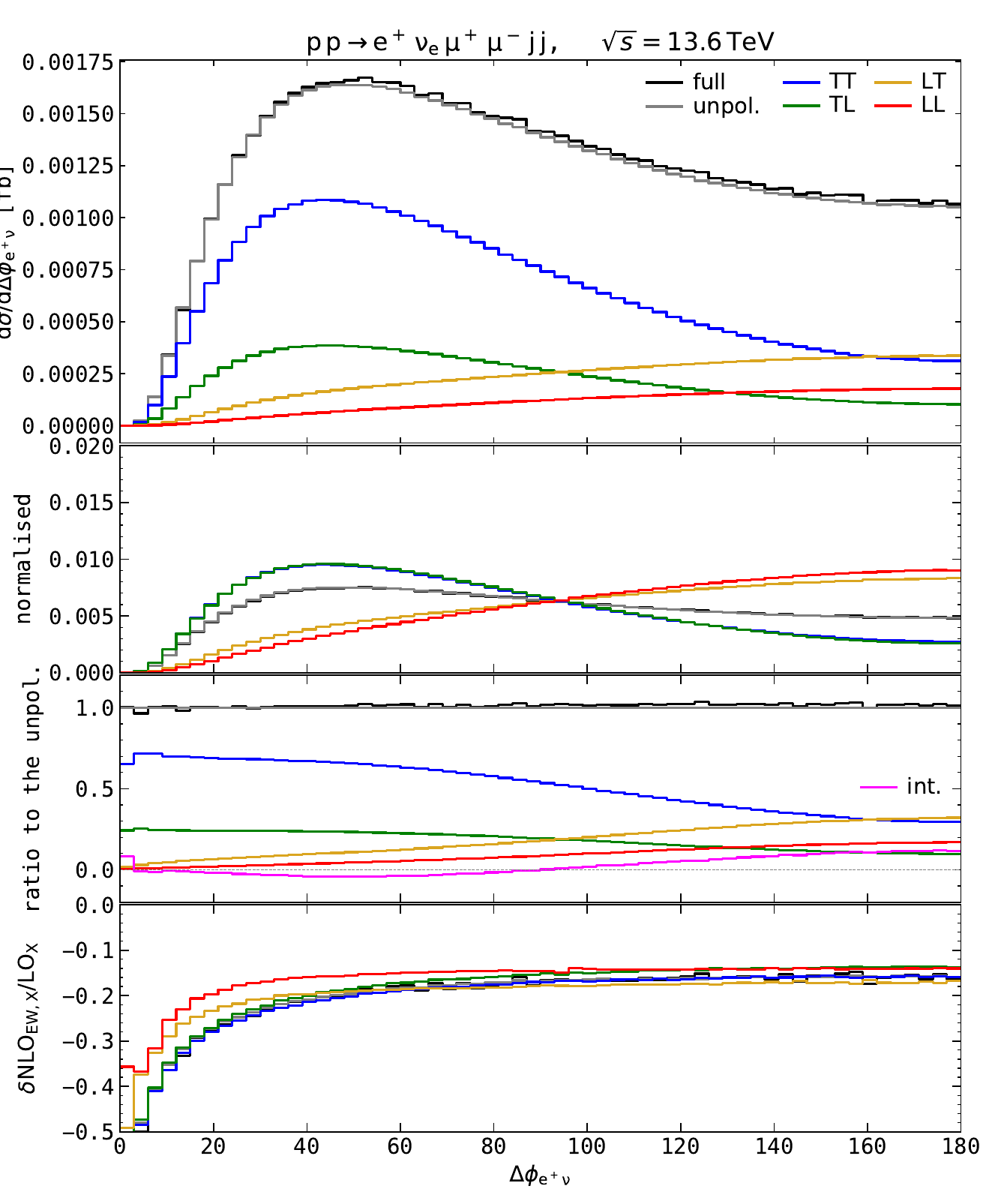}}\\
  \caption{
    Distributions in the cosine of the reconstructed positron decay angle in the $\PW$-boson rest frame (left)
    and in the azimuthal-angle separation between the positron and the neutrino (right)
    for $\PW^+\PZ$ scattering at the LHC in the fiducial setup described in~\refse{sec:ATLASfid}.
    The $\PW$-boson kinematics is obtained through the
    neutrino-reconstruction algorithm of \citere{Aaboud:2019gxl}.
    Same structure as in \reffi{fig:first}.
  }\label{fig:3rd}
\end{figure}

In \reffi{fig:CTSTAR} we consider the distribution in the cosine of the positron decay angle in the
$\PW^+$-boson rest frame relative to the $\PW$-boson spatial direction in the di-boson
CM frame.
To evaluate this observable, the reconstruction of the neutrino kinematics enters
the determination of both the di-boson system and the $\PW$-boson one, therefore leading to a distorted shape
compared to the one obtained using the Monte Carlo-truth neutrino momentum.
This variable features the most direct connection to the polarisation state
of the $\PW$ boson, as shown by the normalised shapes, which are almost identical when comparing the $\rL\rL$ with $\rL\rT$ curve, as well as the $\rT\rT$ and the $\rT\rL$ one.
The presence of cuts and the neutrino reconstruction induces non-vanishing interference terms, which, however, remain at the percent level over the whole angular range.
The NLO EW corrections are rather flat and negative for $\cos\theta^{*}_{\Pe^+,\rm reco}>0$, while they diminish in size for the $\rL\rT$ and $\rL\rL$ signals in the suppressed anti-collinear region.

A very good proxy for $\cos\theta^{*}_{\Pe^+,\rm reco}$, which does not require neutrino reconstruction, is the
azimuthal separation between the positron and the neutrino, evaluated
with (missing) momenta in the laboratory frame.
The distributions for this observable are shown in \reffi{fig:DPHILV}.
The correlation with the decay angle is very strong, just like the high sensitivity
to the $\PW$-boson polarisation state. The similarity between the $\rT\rT$ and $\rT\rL$ normalised shapes is
even stronger than in \reffi{fig:CTSTAR}. The interference effects
rise for 
$\Delta\phi_{\Pe^+\nu}>\pi/2$, and eventually contribute up to $10\%$ of the unpolarised signal.
In the same region, and especially near $\Delta\phi_{\Pe^+\nu}\approx \pi$,
the longitudinal mode for the $\PW$ boson becomes of the same size as the transverse one.
The NLO EW corrections are flat in most of the angular range, with an increase (with negative sign) towards the
suppressed region $\Delta\phi_{\Pe^+\nu}\approx 0$, which is
correlated to highly boosted \PW~bosons.
    
\begin{figure}
  \centering
  \subfigure[\label{fig:DPHIWZ}]{\includegraphics[width=0.48\textwidth, page=6]{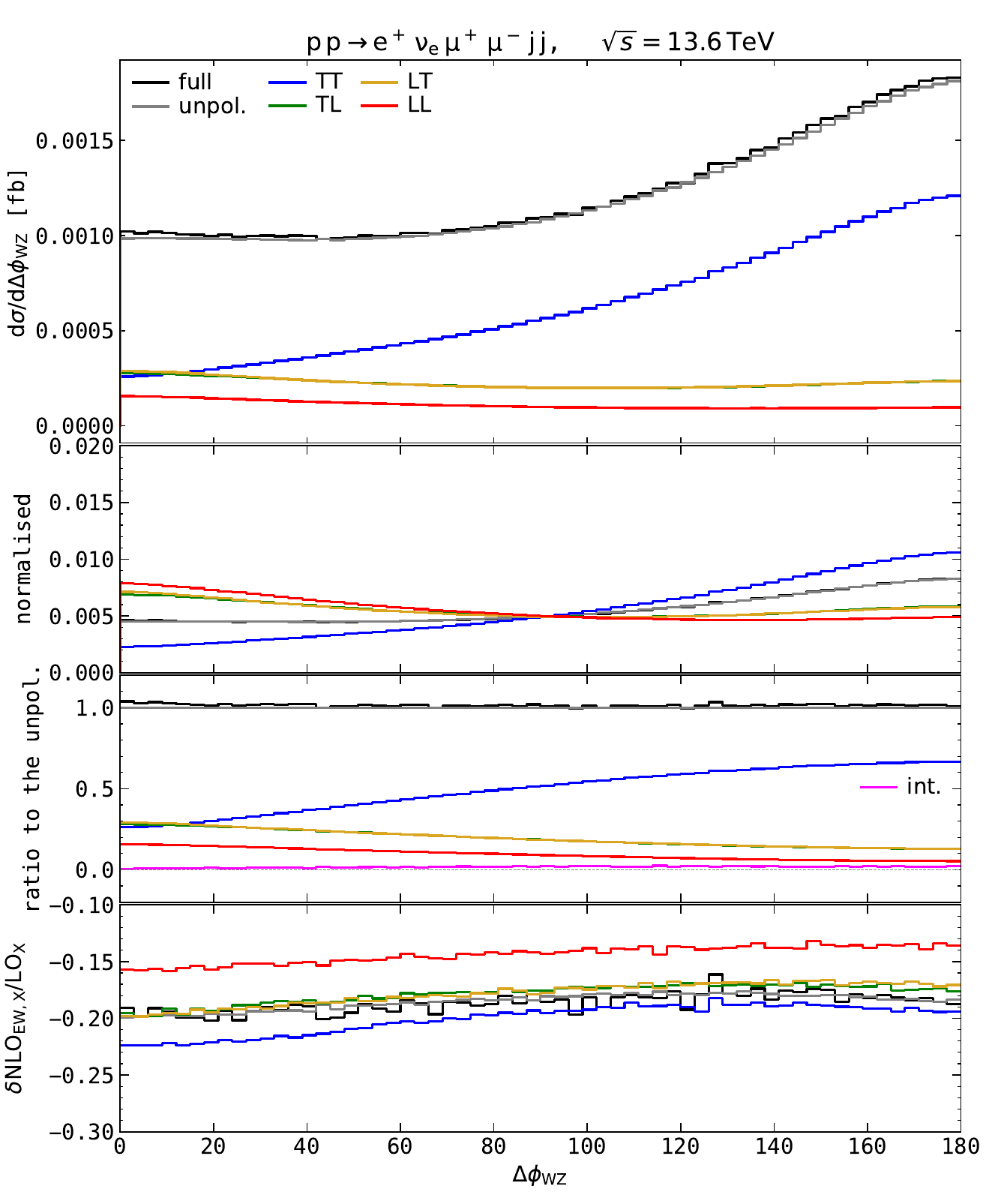}}
  \subfigure[\label{fig:APTJ}]{\includegraphics[width=0.48\textwidth, page=10]{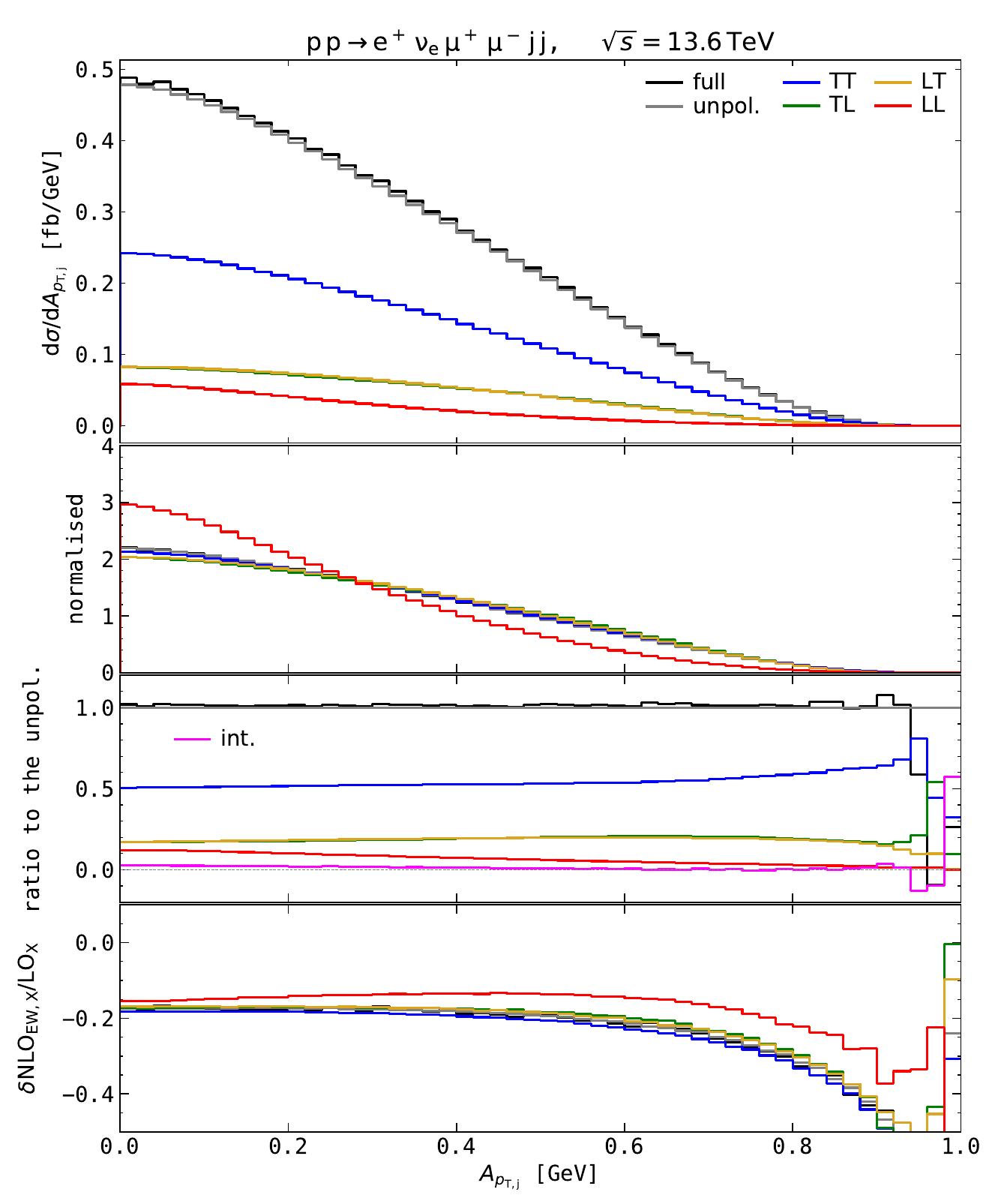}}\\
  \caption{
    Distributions in the azimuthal-angle separation between
    the positron--neutrino and the muon--antimuon systems (left) and the $A_{\pt{\Pj}}$ asymmetry (right) 
    for $\PW^+\PZ$ scattering at the LHC in the fiducial setup described in~\refse{sec:ATLASfid}.
    The definition of $A_{\pt{\Pj}}$ is given in \refeq{eq:Apt_def}.
    Same structure as in \reffi{fig:first}.
  }\label{fig:4th}
\end{figure}

In \reffi{fig:DPHIWZ} a distribution in the azimuthal-angle separation between the
two bosons, \ie between the positron--neutrino and the muon--antimuon
systems, is considered.
This quantity shows a marked shape difference between the $\rT\rT$ curve and the other ones.
In the left endpoint of the angular shape the $\rT\rT$ signal is smaller than the sum of other modes,
while it gets dominant towards the regime where the two bosons are produced in opposite hemispheres in the transverse plane.
At variance with the azimuthal correlation in \reffi{fig:DPHILV}, $\Delta\phi_{\PW\PZ}$ is not
hampered by large interference effects, and also off-shell effects are rather flat and equal to those found at integrated level.
The relative NLO EW corrections are quite flat as well, ranging
between $-13\%$ and $-16\%$ for $\rL\rL$, and between $-19\%$ and $-23\%$ for $\rT\rT$.

In \reffi{fig:APTJ} we study the distribution in the transverse-momentum asymmetry between the two tagging jets, defined as,
\beq \label{eq:Apt_def}
A_{\pt{\Pj}} = \frac{\pt{\Pj_1}-\pt{\Pj_2}}{\pt{\Pj_1}+\pt{\Pj_2}}\,,
\eeq
where $\Pj_{1(2)}$ is the leading (sub-leading) tagging jet, according to transverse-momentum ordering.
Although expected to mostly discriminate between the VBS EW signal and its QCD-induced multi-jet
background, this observable turns out to be also sensitive to polarisation discrimination.
In particular, the $\rL\rL$ signal tends to favour a kinematic regime where the two tagging jets have similar transverse momenta 
in contrast to other polarisation modes. The region with one tagging jet being much more boosted than the other is suppressed
for all polarisation states.
While this observable would be more sizeably modified by NLO QCD corrections (and three-jet topologies in particular),
even the NLO EW corrections affect it in a non-trivial manner.
In particular, while being quite flat in the more symmetric region  ($A_{\pt{\Pj}}\lesssim 0.5$), the EW corrections become
increasingly negative in the more asymmetric region, reaching up to $-50\%$ for $A_{\pt{\Pj}}\approx 0.9$, owing to
large EW Sudakov logarithms.

\begin{figure}
  \centering
  \subfigure[\label{fig:RPT21}]{\includegraphics[width=0.48\textwidth, page=8]{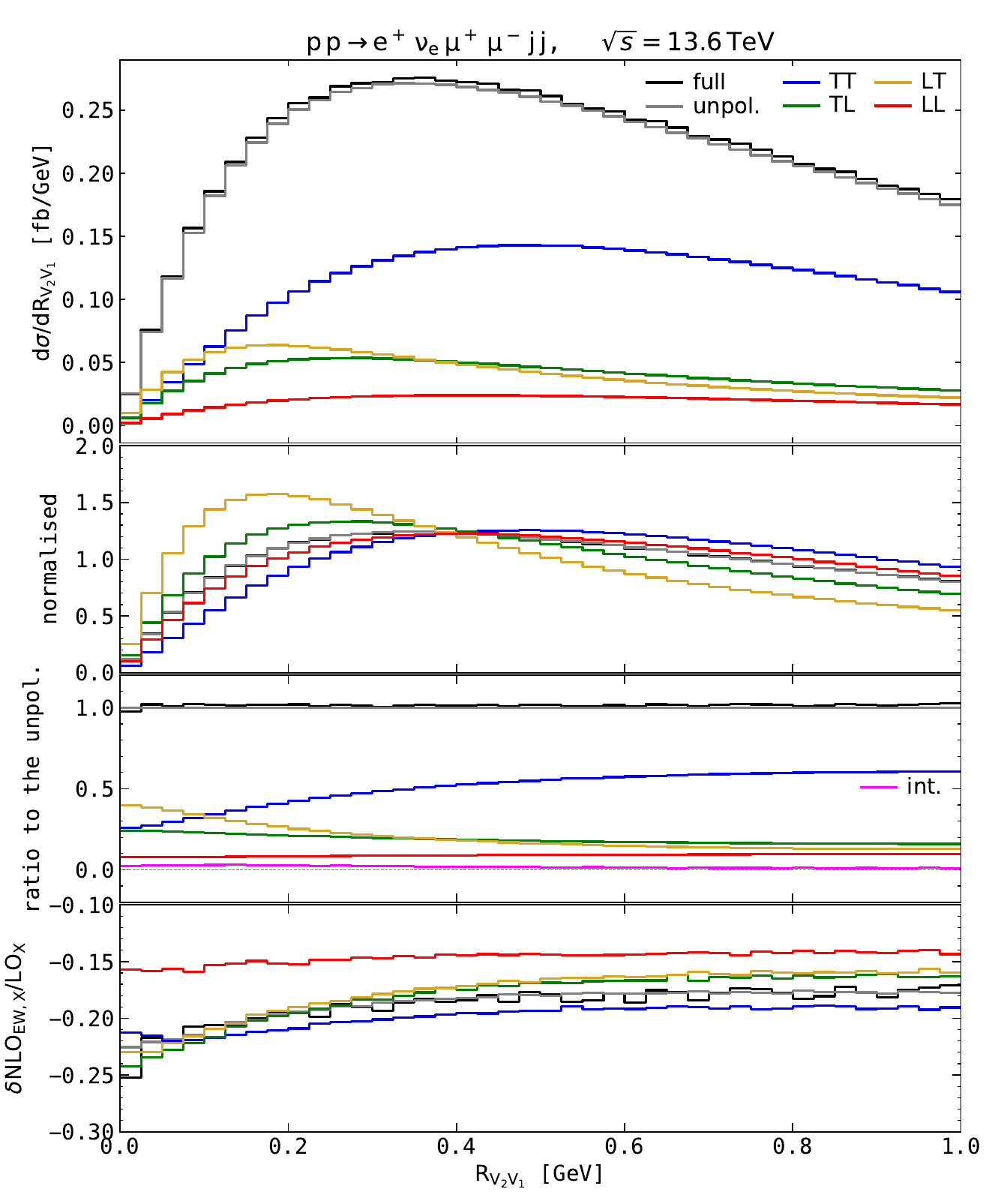}}
  \subfigure[\label{fig:DYEPZ}]{\includegraphics[width=0.48\textwidth, page=4]{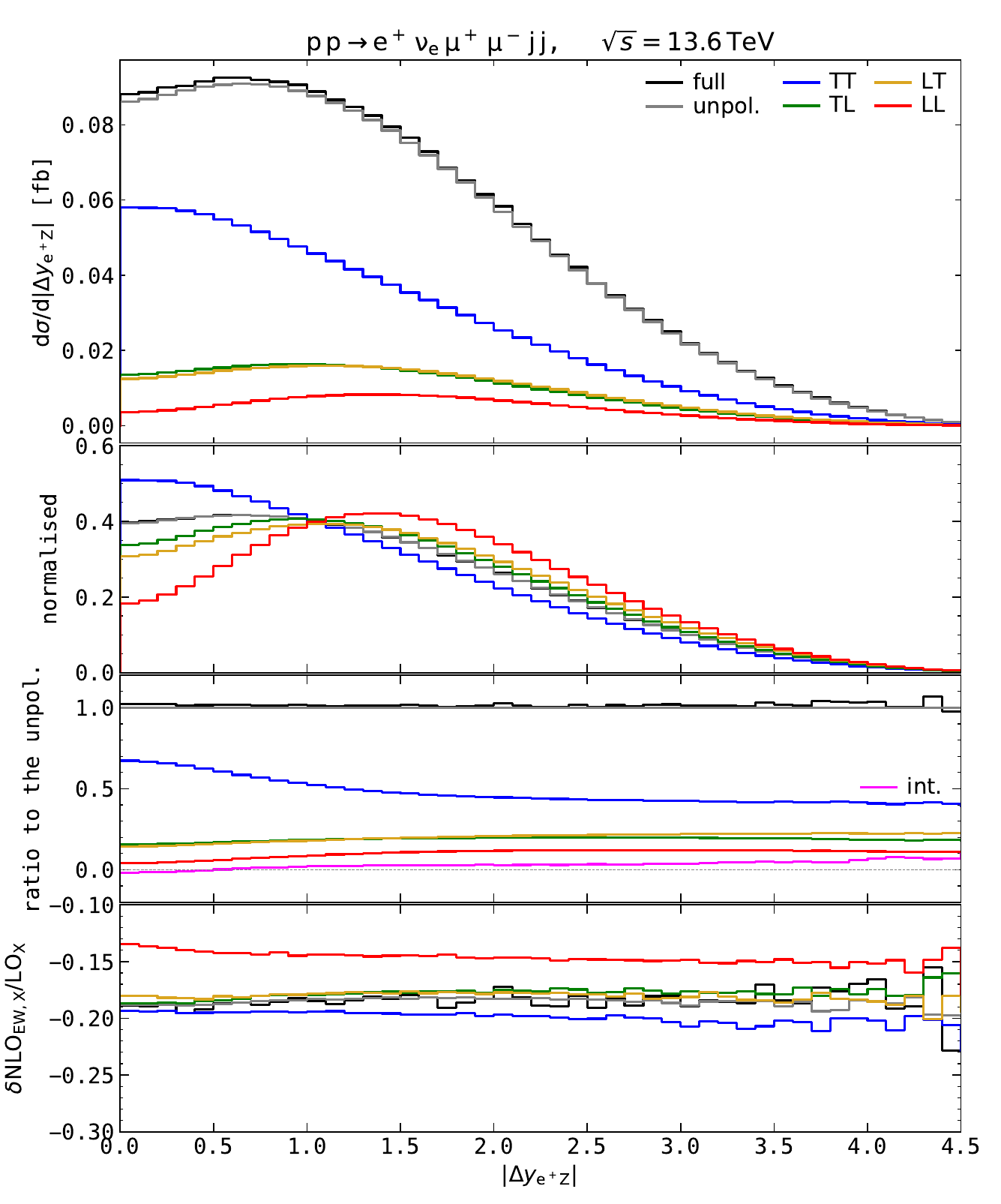}}\\
  \caption{
    Distributions in the the $R_{V_2V_1}$ ratio as defined in \refeq{eq:R21_def} (left)
    and the absolute rapidity separation between the positron and the muon--antimuon system (right) 
    for $\PW^+\PZ$ scattering at the LHC in the fiducial setup described in~\refse{sec:ATLASfid}.
    Same structure as in \reffi{fig:first}.
  }\label{fig:fifth}
\end{figure}

In \reffi{fig:RPT21} we present the distribution in  the
transverse-momentum ratio between the sub-leading and the leading EW
boson, defined as
\beq \label{eq:R21_def}
R_{V_2V_1} = \frac{\pt{V_2}}{\pt{V_1}}\,,
\eeq
where $V_{1(2)}$ is the leading (sub-leading) boson, according to transverse-momentum ordering.
At variance with $\PW^\pm\PW^\pm$ scattering, where the presence of two neutrinos forces to construct
proxies of $R_{V_2V_1}$ with charged-lepton kinematics \cite{Denner:2024tlu},
the presence of a single neutrino in $\PW\PZ$ scattering makes this observable
directly accessible at the LHC.
The normalised shapes are quite different for all polarised signals, with the $\rL\rT$ mode favouring lower
values of $R_{V_2V_1}$ than the other ones. This can be understood with the general tendency that longitudinal
bosons are typically softer than transverse ones. This leads to
a $\rL\rT$ signal being even larger than the
$\rT\rT$ one for $R_{V_2V_1}\lesssim 0.1$.
Compared to the corresponding leptonic proxy introduced in \citere{Denner:2024tlu}, the NLO EW corrections to $R_{V_2V_1}$
are much flatter, especially for the $\rL\rL$ and $\rT\rT$ states. For
the mixed polarisation states, the
EW corrections diminish from $-16\%$ to $-24\%$ towards the least populated kinematic regime. 

As a last differential distribution, in \reffi{fig:DYEPZ} we show the rapidity separation between the positron (from
the $\PW$-boson decay) and the $\PZ$ boson. This variable has been already used in polarisation-oriented
analyses of inclusive $\PW\PZ$ production \cite{ATLAS:2022oge,ATLAS:2024qbd}.
Strikingly, the normalised shapes in VBS show an opposite behaviour
compared to those found in inclusive production
\cite{Le:2022ppa}, 
especially for what concerns the $\rL\rL$ and $\rT\rT$ modes.
In VBS, the $\rL\rL$ curve favours a regime where the positron and the $\PZ$ boson are separated in rapidity
by about $\Delta y_{\Pe^+\PZ}\approx 1.4$, while in inclusive production $\Delta y_{\Pe^+\PZ}\approx 0$ is
preferred. The behaviour of the $\rT\rT$ mode is the opposite,
favouring zero rapidity separation in VBS but larger
values in inclusive production. This difference clearly comes from the different kinematics of the di-boson system
determined by the presence of two energetic jets well separated in rapidity in VBS.
Also the pattern of EW corrections is pretty different compared to inclusive production, with
almost flat corrections relative to the LO results.


\section{Conclusions}\label{sec:conc}
With this work we have accomplished a further step towards the sound interpretation of LHC data in
terms of the polarisation structure associated with the production of EW-boson pairs.
Specifically, we have achieved for the first time exact NLO EW accuracy
in the production and subsequent leptonic decays at the LHC
of one polarised $\PW$ and one polarised $\PZ$ boson in association with two hard jets
in a VBS fiducial phase space.

The calculation relies on the recently developed machinery to simulate polarised EW bosons
as intermediate states in LHC processes, including higher-order effects both in the production
and in the decays. The polarised-boson signal definition is achieved by selecting individual
helicity states in the SM tree-level and one-loop amplitudes, which are treated
in the pole approximation to ensure gauge invariance.

Owing to the physical overlap between VBS and triple-boson production, we devise a generic
treatment of the latter contribution in the presence of higher-order corrections. In this approach,
large weights that potentially hamper the Monte Carlo-integration convergence are removed, while preserving
gauge invariance and
the correct IR cancellation between real-radiation contributions and subtraction counterterms.

A broad phenomenological analysis is carried out in an ATLAS-inspired fiducial volume for LHC collisions
at 13.6\TeV (Run 3), with the $\PW^+\PZ$ pair being produced
in association with two energetic jets well separated in rapidity
and forming a system with a large invariant mass.

The fiducial cross sections show a clear dominance
of the $\rT\rT$ state, while the $\rL\rL$ one contributes less than $10\%$ to the unpolarised result.
The off-shell effects, regarded in our analysis as an irreducible background, are at the $1.5\%$ level,
with slight enhancements only in certain suppressed kinematic configurations. The interference effects
are of the same size at integrated level, while somewhat sizeable
enhancements show up in the spectrum of
certain azimuthal-angle distributions or at low transverse mass of the di-boson system.

The high-energy behaviour (large invariant masses or transverse momenta)
depends on the polarisation state, with the $\rL\rL$ distributions
typically falling off faster than mixed and purely transverse ones. In the same regions
the asymptotic pattern of relative NLO EW corrections confirms the dominance of
leading EW Sudakov logarithms, with the negative corrections for longitudinal bosons being smaller
in size than for the transverse ones owing to smaller EW Casimir factors.
Even at the integrated level the NLO EW corrections are very large and negative, ranging between $-14\%\,(\rL\rL)$ and $-20\%\,(\rT\rT)$.

From the investigation of angular observables, it turns out that the
NLO EW corrections are rather flat in the most populated regions, and the normalised shapes
point to a marked sensitivity to the longitudinal mode of one (\eg $\cos\theta^{*}_{\Pe^+}$) or two bosons
(\eg $\Delta y_{\Pe^+\PZ}$).
Owing to the tendency of longitudinal bosons to be softer (in transverse momentum) than the transverse ones,
a number of observables associated to the tagging-jet kinematics ($\pt{\Pj_1\Pj_2},\,A_{\pt{\Pj}}$)
has been found to be also fairly sensitive to the $\rL\rL$ mode.

Overall, merging the integrated and differential results of this work to construct suitable boosted-decision trees or neural-network architectures could help further maximising the potential of discrimination between longitudinal and transverse modes
in $\PW\PZ$ scattering (and other VBS channels as well).

We conclude with a remark on QCD corrections.
We have shown for $\PW^+\PW^+$ scattering that the NLO QCD corrections are more
independent of polarisation states than the NLO EW ones.
Therefore, we expect that applying (differential) QCD-correction factors from off-shell calculations
to our NLO EW results for polarised bosons
is a decent approximation, provided that a corresponding systematic uncertainty is accounted for
in data analyses.
The actual calculation of NLO QCD corrections, or better said of  $\mc O(\alphas\alpha^6)$,
to polarised $\PW\PZ$ scattering is left for future work.

\section*{Acknowledgements}
The authors thank Olivier Arnaez, Angela Maria Burger, Alberto
Carnelli, Lucia Di Ciaccio, Mathis Dubau, Iro Koletsou,
Santiago Lopez Portillo Chavez, Mathieu Pellen,
and Emmanuel Sauvan for fruitful discussions.

AD and RF are supported by the German Federal Ministery for Education
and Research (BMBF) under contract no.~05H24WWA.
AD and DL are supported by the German Research Foundation (DFG) under
reference number DE 623/8-1.
The research of DL has also been partially supported by
the Italian Ministry of Universities and Research (MUR) under the FIS grant (CUP: D53C24005480001, FLAME).
GP acknowledges support from 
the EU Horizon Europe research and innovation programme
under the Marie-Sk\l{}odowska Curie Action
``POEBLITA - POlarised Electroweak Bosons at the LHC with Improved Theoretical Accuracy'' - grant agreement no.~101149251 (CUP H45E2300129000)
and the Italian MUR, with EU funds (NextGenerationEU), through the
PRIN2022 grant agreement no.~20229KEFAM (CUP H53D23000980006).

\bibliographystyle{JHEP} 
\bibliography{pol}

\end{document}